\documentclass[a4paper,12pt]{iopart}

\usepackage[USenglish]{babel}
\usepackage{amssymb}
\usepackage{hyperref}
\usepackage{graphicx}
\usepackage{color}
\usepackage[utf8]{inputenc}
\usepackage{bm}
\usepackage{cite}

\newcommand{\moire}{moir\'e }
\newcommand{\Moire}{Moir\'e }
\newcommand{\bsymb}[1]{{\bm{#1}}}
\newcommand{\conj}[1]{#1^*}

\newcommand{\ind}[3]{{#1}_{#2}^{(#3)}}
\newcommand{\erw}[1]{\left\langle #1 \right\rangle}

\newcommand{\pdag}{\vphantom{\dagger}}

\newcommand{\comm}[2]{\left[#1, #2 \right]}
\newcommand{\dd}[1]{{\rm d} #1}
\newcommand{\dv}[1]{\frac{{\rm d}}{{\rm d} #1}}
\newcommand{\pv}{\mathcal{P}}
\renewcommand{\Re}{{\rm Re}}
\renewcommand{\Im}{{\rm Im}}
\newcommand{\abs}[1]{\left| #1 \right|}
\newcommand{\ket}[1]{\left| #1 \right>}

\begin{document}

\title{Theory of phonon sidebands in the absorption spectra of \moire exciton-polaritons}

\author{K.~J\"urgens}
\address{Institute of Solid State Theory, University of M\"unster, 48149 M\"unster, Germany}
\author{D.~Wigger}
\address{Department of Physics, University of M\"unster, 48149 M\"unster, Germany}
\author{T.~Kuhn}
\address{Institute of Solid State Theory, University of M\"unster, 48149 M\"unster, Germany}

\begin{abstract}
Excitons in twisted bilayers of transition metal dichalcogenides have strongly modified dispersion relations due to the formation of periodic \moire potentials.
The strong coupling to a light field in an optical cavity leads to the appearance of \moire polaritons.
In this paper, we derive a theoretical model for the linear absorption spectrum of the coupled \moire polariton-phonon system based on the time-convolutionless (TCL) approach.
Results obtained by numerically solving the TCL equation are compared to those obtained in the Markovian limit and from a perturbative treatment of non-Markovian corrections.
A key quantity for the interpretation of the findings is the generalized phonon spectral density.
We discuss the phonon impact on the spectrum for realistic \moire exciton dispersions by varying twist angle and temperature.

Key features introduced by the coupling to phonons are broadenings and energy shifts of the upper and lower polariton peak and the appearance of phonon sidebands between them.
We analyze these features with respect to the role of Markovian and non-Markovian effects and find that they strongly depend on the twist angle.
We can distinguish between the regimes of large, small, and intermediate twist angles.
In the latter phonon effects are particularly pronounced due to dominating phonon transitions into regions which are characterized by van Hove singularities in the density of states.
\end{abstract}

\maketitle

\section{Introduction}

\begin{figure}
    \centering
    \includegraphics[width=0.6\textwidth]{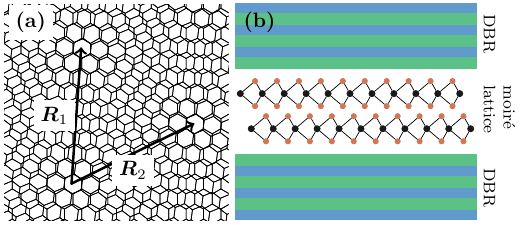}
    \caption{(a) Sketch of two twisted hexagonal lattices with the resulting \moire lattice vectors $\bsymb{R}_{1,2}$. (b) Side view of the \moire bilayer in an optical cavity formed by two DBRs.}
    \label{fig:fig1}
\end{figure}

Twisted heterostructures of van der Waals monolayers are an interesting class of materials with strongly modified electronic \cite{he2021moi,yoo2019ato,wang2020corr}, excitonic \cite{zhang2018moi,villafane2023twi,knorr2022exc,brem2024opt,wu2022evi,wang2021moi,seyler2019sig}, phononic \cite{li2023rev,koshino2019moi,koshino2020eff,lin2018moi,nam2017lat,chuang2022eme} and optical \cite{baek2020highly,wu2018the,wu2022evi,brem2024opt,yu2015ano} properties compared to the respective monolayers.
The relative rotation between two monolayers leads to a modified electronic environment for charge carriers in the heterostructure, which results in the formation of a periodic \moire potential \cite{he2021moi,liu2023exa,shabani2021dee,wu2018the,lin2023rem}.
The periodicity of the twisted heterostructure in real space is sketched in Fig.~\ref{fig:fig1}~(a), where $\bsymb{R}_{1,2}$ are the \moire lattice vectors.
Excitons in twisted monolayers of transition metal dichalcogenide (TMDC) interact with the \moire potential and can be localized at its minima.
At small twist angles the distance between two neighboring minima is large, leading to a negligible interaction between the localized excitons in adjacent minima and, correspondingly, a flat exciton dispersion \cite{he2021moi,brem2020tun,gotting2022moi,knorr2022exc}.
Such trapped excitons have been shown to feature single photon emission \cite{baek2020highly,yu2017moi}.
This situation is often compared to an array of periodically aligned quantum dots \cite{baek2020highly,lohof2023con,wu2018the}.
For larger twist angles and, thus, shorter distances to the neighboring minima the interaction between excitons in different minima increases leading to delocalized exciton states and a more 2D-like behavior \cite{brem2020tun,knorr2022exc,gotting2022moi}.
In addition to the modification of the electronic states, the \moire potential also impacts the crystal lattice, leading to the formation of \moire phonon modes \cite{koshino2019moi,lin2018moi} and a modified exciton-phonon coupling \cite{koshino2020eff,shinokita2021res,lim2023mod}.

When integrating the twisted heterostructure in a photonic cavity formed, for example, by two distributed Bragg reflectors (DBRs), as sketched in Fig.~\ref{fig:fig1}~(b), the \moire excitons interact with photons of the cavity leading to the formation of \moire exciton-polaritons \cite{fitzgerald2022twi,zhang2021van,forg2019cav}.
Due to their low effective mass they are interesting candidates for Bose-Einstein condensation \cite{kasprzak2006bos,kessler2008lig}.

In this contribution we investigate the phonon impact on the optical properties of \moire exciton-polaritons for varying twist angles and temperatures.
After the introduction of the model (Sec.~\ref{sec:hamiltonian}) and the optical absorption spectrum (Sec.~\ref{sec:spectrum}), in Sec.~\ref{sec:eom} a correlation expansion up to second order Born approximation \cite{rossi2002the,lengers2020the} together with a time-convolutionless (TCL) approach \cite{lengers2020the,lengers2021phon} are used to derive a closed set of coupled differential equations for the dynamics of the polariton polarization with time-dependent coefficients.
This approach has been successfully applied to excitons and polaritons in single TMDC layers \cite{lengers2020the,lengers2021phon}, and to the calculation of nonlinear optical signals \cite{richter2010tim}.
Here, it will be extended to polaritons in twisted bilayers.
In Sec.~\ref{sec:assumptions} the parameters chosen for our calculations are specified.

In Sec.~\ref{sec:real_spec} the TCL equations are solved numerically for a realistic \moire exciton dispersion.
This allows us to discuss the optical spectra for varying twist angles and temperatures.
In Sec.~\ref{sec:comp_pert} the numerically calculated results are compared with a perturbative solution which helps us in obtaining a better understanding of the features in the spectra and in particular of the significance of (non-)Markovian processes.
In Sec.~\ref{sec:twist_dep} we then analyze the twist angle dependence of the key features of the spectra, in particular the convergence of the perturbative approach as well as the peak positions and widths.
Interestingly, the phonon coupling leads to a clear separation of regimes for small and large twist angles, separated by a transition region where the phonon influence is particularly pronounced.
Finally, in Sec.~\ref{sec:conclus} we summarize our results with some concluding remarks.

\section{Theory}

\subsection{Hamiltonian}
\label{sec:hamiltonian}We consider excitons in a twisted TMDC bilayer that is embedded in a photonic cavity as sketched in Fig.~\ref{fig:fig1}~(b).
Due to the twist between the monolayers by an angle $\vartheta$ a two-dimensional periodic structure forms, characterized by two \moire lattice vectors $\bsymb{R}_1$ and $\bsymb{R}_2$, as schematically plotted in Fig.~\ref{fig:fig1}~(a).
The periodic structure leads to a periodic \moire potential for the excitons, which can be written as
\begin{equation}
    V(\bsymb{r}) = \sum_{j} \tilde{V}_{\bsymb{G}_j} e^{i\bsymb{G}_j \cdot \bsymb{r}},
    \label{eq:moire_potential}
\end{equation}
where $\bsymb{G}_j$ is the $j$th reciprocal \moire lattice vector and $\tilde{V}_{\bsymb{G}_j}$ is the corresponding Fourier component of the \moire potential.
In the following we will assume equal lattice constants of the monolayers.
Then, for all twist angles except for multiples of $60^\circ$ the periodicity of the \moire potential is larger than the periodicity of the monolayers, leading to a 1st \moire Brillouin zone (MBZ)  $\mathcal{B}_{\rm{m}}$ which is smaller than the 1st Brillouin zone of the monolayers.

This periodic \moire potential leads to the formation of \moire excitons \cite{zhang2018moi,villafane2023twi,wu2022evi,brem2020tun,knorr2022exc}, described by the creation (annihilation) operators $X_{n, \bsymb{k}}^\dagger$ ($X_{n, \bsymb{k}}^{\pdag}$) with band index $n$, center of mass wave vector $\bsymb{k}$ with $\bsymb{k}\in \mathcal{B}_{\rm m}$, and dispersion relation $\hbar \omega_{n,\bsymb{k}}$.
The dispersion strongly depends on the twist angle $\vartheta$ between the two layers; it is flat for small $\vartheta$ \cite{gotting2022moi,brem2020tun} and becomes more 2D-like with increasing $\vartheta$ due to the decreasing period length of the \moire potential in real space.
Examples of exciton dispersions for different values of $\vartheta$ can be found in Fig.~\ref{app:fig8} in the \ref{app:sec_moire-excitons}.

The derivation of the \moire exciton operators and their dispersion relation starting from the electron-hole picture for a homogeneous bilayer without \moire potential is given in \ref{app:sec_moire-excitons}.
We basically follow the derivation in Refs.~\cite{brem2020tun,knorr2022exc}.
In a first step, operators for excitons of the homogeneous bilayer are introduced by solving a Wannier equation incorporating the electron-hole interaction but without a \moire potential.
In a second step, the \moire excitons are introduced by diagonalizing the Hamiltonian of the homogeneous excitons extended by the \moire potential.
This involves a back-folding of the homogeneous exciton bands into the 1st MBZ.
Note, that in this paper we will use the expressions \textit{homogeneous exciton} and \textit{\moire exciton} to distinguish between the excitons of the 2D bilayer \textit{without} and \textit{with} the impact of the \moire potential, respectively.

The \moire excitons interact with the phonon modes of the heterostructure.
The phonons are described by the creation (annihilation) operators $b_{j, \bsymb{Q}}^\dagger$ ($b_{j, \bsymb{Q}}^{\pdag}$) for a phonon in the branch $j$ with wave vector $\bsymb{Q}$ and dispersion relation $\hbar \Omega_{j, \bsymb{Q}}$.
We neglect the influence of the \moire lattice on the phonons, therefore the phonon wave vector $\bsymb{Q}$ is not restricted to the 1st MBZ.
The phonons couple to the \moire excitons via the coupling constant $\hbar \mathcal{G}_{j, \bsymb{k}, \bsymb{Q}}^{(n, n')}$.
In \ref{app:sec_phonon_coupling} this coupling constant is derived starting from the electron/hole-phonon coupling Hamiltonian and inserting the definition of the \moire exciton operators from \ref{app:sec_moire-excitons}.

Embedding the \moire heterostructure in a photonic cavity finally leads to a coupling of the \moire excitons to cavity photons.
The photons are described by the creation (annihilation) operators $a_\bsymb{K}^\dagger$ ($a_\bsymb{K}^{\pdag}$) for a photon in the considered cavity mode with in-plane wave vector $\bsymb{K}$.
We restrict the photons to the subspace of a single mode in the out-of-plane direction with discrete wave vector $K_z$ and energy $\hbar \omega_\bsymb{0} = \hbar c K_z$, $c$ being the velocity of light in the medium, and a continuum of in-plane modes with wave vectors $\bsymb{K}$, leading to the dispersion relation 
\begin{eqnarray}
    \hbar \omega_{\bsymb{K}} &= \hbar c \sqrt{\bsymb{K}^2 + K_z^2} \nonumber \\
        &= \hbar \omega_\bsymb{0} \sqrt{1 + \left(\frac{c\bsymb{K}}{\omega_\bsymb{0}}\right)^2}.
        \label{eq:photon}
\end{eqnarray}
The in-plane photon wave vectors are not restricted to the 1st MBZ, but for each $\bsymb{K}$ one can find a reciprocal lattice vector $\bsymb{G}_m$, such that $\bsymb{K} = \bsymb{k} + \bsymb{G}_m$ with $\bsymb{k}\in \mathcal{B}_{\rm m}$.
In general, \moire excitons with wave vector $\bsymb{k}$ couple to all photon modes $\bsymb{K}$ that differ only by a reciprocal lattice vector $\bsymb{G}_m$.
However, assuming, that the 1st MBZ is sufficiently large, the photons from a neighboring MBZ with $\bsymb{K} =\bsymb{k} + \bsymb{G}_m$ and $\bsymb{G}_m \neq \bsymb{0}$ are effectively decoupled from the excitons in the 1st MBZ due to the steep photon dispersion.
Therefore, the coupling between exciton and photon modes that differ by a reciprocal lattice vector $\bsymb{G}_m \neq \bsymb{0}$ can usually be neglected and it is sufficient to take into account photon modes with $\bsymb{k}\in \mathcal{B}_{\rm m}$.
The exciton-photon coupling is then described by the coupling strength $\lambda_{n, \bsymb{k}}$.
In systems where this is not fulfilled or where the cavity field is resonant to excitons close to the edge of the 1st MBZ, a coupling of multiple photon modes to the exciton via photon-umklapp processes may occur, as discussed in Ref.~\cite{kessler2008lig}.
Here, however, we will not consider this case.

The total Hamiltonian for the exciton-photon-phonon system then reads
\begin{eqnarray}
    H &= \sum_{n, \bsymb{k}} \hbar \omega_{n, \bsymb{k}}^{\pdag} X_{n, \bsymb{k}}^\dagger X_{n, \bsymb{k}}^{\pdag} + \sum_{\bsymb{k}} \hbar \omega_{\bsymb{k}}^{\pdag} a_{\bsymb{k}}^\dagger a_{\bsymb{k}}^{\pdag} \nonumber \\
    &+ \sum_{n, \bsymb{k}} \hbar \left[\lambda_{n, \bsymb{k}}^{*} a_{\bsymb{k}}^\dagger X_{n, \bsymb{k}}^{\pdag} + \lambda_{n, \bsymb{k}}^{\pdag} a_{\bsymb{k}}^{\pdag} X_{n, \bsymb{k}}^\dagger \right] \nonumber \\
    &+ \sum_{j, \bsymb{Q}} \hbar \Omega_{j, \bsymb{Q}}^{\pdag} b_{j, \bsymb{Q}}^\dagger b_{j, \bsymb{Q}}^{\pdag}  \nonumber \\
    &+ \sum_{{n, n', j, \bsymb{k}, \bsymb{Q}, \bsymb{G}}} \hbar \mathcal{G}_{j, \bsymb{k}, \bsymb{Q}}^{(n, n')} X_{n', \bsymb{k} + \bsymb{Q} + \bsymb{G}}^\dagger X_{n, \bsymb{k}}^{\pdag} \left( b_{j, \bsymb{Q}}^{\pdag} + b_{j, -\bsymb{Q}}^\dagger \right) 
    \label{eq:H_bloch}
\end{eqnarray}
where $\bsymb{k}\in \mathcal{B}_{\rm m}$ and $\bsymb{G}$ is chosen such that $\bsymb{k} + \bsymb{Q} + \bsymb{G} \in \mathcal{B}_{\rm m}$.

The exciton-photon part, i.e., the first three sums in Eq.~\ref{eq:H_bloch}, can be diagonalized by introducing the polariton operators
\begin{eqnarray}
    \label{eq:def_P}
    P_{\Lambda, \bsymb{k}} &= \beta_{\bsymb{k}; \Lambda, 0} a_{\bsymb{k}} + \sum_{n=1}^N \beta_{\bsymb{k}; \Lambda, n} X_{n, \bsymb{k}}
\end{eqnarray}
with the normalization condition $\sum_\Lambda \conj{\beta_{\bsymb{k}; \Lambda, n}} \beta_{\bsymb{k}, \Lambda, n'}^{\pdag} = \delta_{n, n'}$, the number of \moire exciton modes $N$, and the quantum number for the polariton branch $\Lambda$ with $\Lambda = 0,1,\dots,N$.
Exciton and photon operators are obtained from the polariton operators by the inverse transformation
\begin{eqnarray}    
    \label{eq:def_P_inv}
    \sum_\Lambda \conj{\beta_{\bsymb{k}; \Lambda, l}} P_{\Lambda, \bsymb{k}} &= \delta_{l, 0} a_\bsymb{k} + \sum_{n=1}^N \delta_{l,n} X_{n, \bsymb{k}}.
\end{eqnarray}
For small excitation densities the exciton operators can be treated as bosonic operators with $\comm{X_{n, \bsymb{k}}^{\pdag}}{X_{n', \bsymb{k}'}^\dagger} \approx \delta_{n,n'} \delta_{\bsymb{k}, \bsymb{k}'}$.
This translates to a bosonic commutation relation for the polariton operators
\begin{eqnarray}
    \comm{P_{\Lambda, \bsymb{k}}^{\pdag}}{P_{\Lambda', \bsymb{k}'}^{\dagger}} \approx \delta_{\Lambda, \Lambda'}^{\pdag} \delta_{\bsymb{k}, \bsymb{k}'}^{\pdag}.
\end{eqnarray}

The Hamiltonian in this basis then reads
\begin{eqnarray}
    H &= \sum_{\Lambda, \bsymb{k}} \hbar \epsilon_{\Lambda, \bsymb{k}}^{\pdag} P_{\Lambda, \bsymb{k}}^\dagger P_{\Lambda, \bsymb{k}}^{\pdag} + \sum_{j, \bsymb{Q}} \hbar \Omega_{j, \bsymb{Q}} b_{j, \bsymb{Q}}^\dagger b_{j, \bsymb{Q}} \nonumber \\
    &+ \sum_{{\Lambda, \Lambda', j, \bsymb{k}, \bsymb{Q}, \bsymb{G}}} 
    \hbar \ind{g}{j, \bsymb{k}, \bsymb{Q}}{\Lambda, \Lambda'}
    P_{\Lambda', \bsymb{k} + \bsymb{Q} + \bsymb{G}}^\dagger P_{\Lambda, \bsymb{k}}^{\pdag}
    \left( b_{j, \bsymb{Q}}^{\pdag} + b_{j, -\bsymb{Q}}^\dagger \right)
    \label{eq:H_pol}
\end{eqnarray}
with the polariton dispersion relation $\hbar \epsilon_{\Lambda, \bsymb{k}}$, that can be calculated from the exciton and photon dispersion relations.
The polariton-phonon coupling reads 
\begin{eqnarray}
    \ind{g}{j, \bsymb{k}, \bsymb{Q}}{\Lambda, \Lambda'} &= \sum_\bsymb{G} \sum_{n, n'=1}^N \mathcal{G}_{j, \bsymb{k}, \bsymb{Q}}^{(n, n')} \conj{\beta_{\bsymb{k}; \Lambda, n}} \beta_{\bsymb{k} + \bsymb{Q} + \bsymb{G}; \Lambda', n'}^{\pdag}
\end{eqnarray}
with $\bsymb{G}$ chosen such that $\bsymb{k} + \bsymb{Q} + \bsymb{G} \in \mathcal{B}_{\rm m}$.
As seen in Eq.~\ref{eq:H_pol}, the phonons lead to transitions within one ($\Lambda=\Lambda'$) or between different polariton branches $\Lambda$ and $\Lambda'$.
This could also involve transitions to momentum-dark polariton branches \cite{ferreira2023sig} which, however, will not be considered here.

\subsection{Absorption spectrum}
\label{sec:spectrum}
We are interested in the absorption spectrum of light impinging on the cavity under an angle $\theta$, i.e., light with an in-plane wave vector $\bsymb{k}$ with $k=(\omega/c_0)\sin (\theta)$, $c_0$ being the vacuum speed of light.
The light couples through the top mirror to the cavity photons with the same in-plane wave vector.
The absorption spectrum $\alpha_{\bsymb{k}}(\omega)$ is then obtained from the response of the cavity mode to a delta-pulse like excitation at time $t=0$, which effectively sets an initial value of the cavity field amplitude $\erw{a_{\bsymb{k}}}(t=0)$.
We will measure the photon energy $\omega$ relative to the cavity mode frequency $\omega_\bsymb{0}$ at $\bsymb{k} = \bsymb{0}$.
The spectrum is then given by the real part of the Fourier transform of the field amplitude,
\begin{eqnarray}
    \alpha_{\bsymb{k}}(\omega) &= \Re\left(\int\limits_0^\infty \dd{t} \erw{a_{\bsymb{k}}}(t) e^{i (\omega+\omega_\bsymb{0}) t} \right) \nonumber \\
        &=\Re\left(\sum_{\Lambda} \conj{\beta_{\bsymb{k}; \Lambda, 0}} \int\limits_0^\infty \dd{t} \erw{P_{\Lambda, \bsymb{k}}}(t) e^{i (\omega+\omega_\bsymb{0}) t} \right),
        \label{eq:spectrum}
\end{eqnarray}
where we have used the inverse transformation from the polariton to the photon operator according to Eq.~\ref{eq:def_P_inv}.
In this paper we will concentrate on the spectra $\alpha_{\bsymb{0}}(\omega)$ for normal incidence, i.e., for $\bsymb{k} = \bsymb{0}$.

\subsection{Equations of motion}
\label{sec:eom}
As we have seen in Eq.~\ref{eq:spectrum}, the quantity of interest for the calculation of the absorption spectrum is the expectation value of the polariton operator $P_{\Lambda, \bsymb{k}}$.
Therefore, we now derive the equation of motion for the polariton polarization $\tilde{P}_{\Lambda, \bsymb{k}} = \erw{P_{\Lambda, \bsymb{k}}} e^{i\omega_{\bsymb{0}}t}$, conveniently defined in a frame rotating with the frequency of the light field at $\bsymb{k} = \bsymb{0}$.
Applying Heisenberg's equation of motion to the polariton operator leads to
\begin{eqnarray}
    \label{eq:P1}
    \dv{t} \tilde{P}_{\Lambda, \bsymb{k}} &= 
        -i \Delta_{\Lambda, \bsymb{k}} \tilde{P}_{\Lambda, \bsymb{k}} 
        -i\sum_{j, \Lambda',\bsymb{Q}} 
            \left[ 
                \ind{\tilde{S}}{j, \bsymb{k}, \bsymb{Q}}{\Lambda, \Lambda', -} 
                + \ind{\tilde{S}}{j, \bsymb{k}, \bsymb{Q}}{\Lambda, \Lambda', +}
                \right]
\end{eqnarray}
with the \moire polariton dispersion relative to the photon mode $\Delta_{\Lambda, \bsymb{k}} = \epsilon_{\Lambda, \bsymb{k}} - \omega_{\bsymb{0}}$ and the phonon-assisted polariton polarizations (PAPPs) in the same rotating frame, defined as
\begin{numparts}
\label{eq:PAPP}
\begin{eqnarray}
    \ind{\tilde{S}}{j, \bsymb{k}, \bsymb{Q}}{\Lambda, \Lambda', -} 
        &= \left[\ind{g}{j, \bsymb{k}, -\bsymb{Q}}{\Lambda, \Lambda'} \right]^*
            \sum_\bsymb{G} \erw{P_{\Lambda', \bsymb{k} - \bsymb{Q} - \bsymb{G}}^{\pdag} b_{j, \bsymb{Q}}^{\pdag}}
            e^{i\omega_{\bsymb{0}}t} , \\
    \ind{\tilde{S}}{j, \bsymb{k}, \bsymb{Q}}{\Lambda, \Lambda', +} 
        &= \left[\ind{g}{j, \bsymb{k}, -\bsymb{Q}}{\Lambda, \Lambda'} \right]^*
            \sum_\bsymb{G} \erw{P_{\Lambda', \bsymb{k} - \bsymb{Q} - \bsymb{G}}^{\pdag} b_{j, -\bsymb{Q}}^\dagger} e^{i\omega_{\bsymb{0}}t} ,
\end{eqnarray}
\end{numparts}
where $\bsymb{G}$ is chosen such that $\bsymb{k} - \bsymb{Q} - \bsymb{G} \in \mathcal{B}_{\rm m}$ and where we have used the relation $\left[\ind{g}{j, \bsymb{k}, \bsymb{Q}}{\Lambda, \Lambda'} \right]^* = \sum_\bsymb{G} \ind{g}{j, \bsymb{k + \bsymb{Q} + \bsymb{G}}, -\bsymb{Q}}{\Lambda', \Lambda}$.

When calculating the equation of motion for $\ind{\tilde{S}}{j, \bsymb{k}, \bsymb{Q}}{\Lambda, \Lambda', \pm}$ we find higher-order phonon-assisted expectation values, which finally results in an infinite hierarchy of equations of motion.
This hierarchy can be truncated using a correlation expansion \cite{rossi2002the,lengers2020the}, assuming that correlations among an increasing number of operators become less important.
Due to the assumption of a small number of photons and excitons and since we are interested in the linear response spectrum, we neglect all terms with more than one polariton operator.
We also restrict our calculations to the 2nd order Born approximation \cite{lengers2020the}, i.e., we neglect all terms that are higher than 2nd order in the polariton-phonon coupling.
In combination with the small polariton density this implies that all expectation values of $b_{j, \bsymb{Q}}^{\pdag}$, $b_{j, \bsymb{-Q}}^\dagger$ and all their combinations are constant (except for self oscillations)~\cite{lengers2020the}.
Assuming that the phonons are initially in a thermal state with temperature $T$, the expectation value $\erw{b_{j', \bsymb{Q}'}^\dagger b_{j, \bsymb{Q}}^{\pdag}} = \delta_{j, j'} \delta_{\bsymb{Q}, \bsymb{Q}'} n_{j, \bsymb{Q}}$ is diagonal and proportional to the thermal phonon occupation $n_{j, \bsymb{Q}}$ and $\erw{b_{j, \bsymb{Q}}^{\pdag}} = \erw{b_{j, \bsymb{Q}}^\dagger} = 0$.

The remaining equation of motion for $\ind{\tilde{S}}{j, \bsymb{k}, \bsymb{Q}}{\Lambda, \Lambda', \pm}$ reads
%\begin{widetext}
\begin{eqnarray}
    \label{eq:S+-}
    \dv{t} \ind{\tilde{S}}{j, \bsymb{k}, \bsymb{Q}}{\Lambda, \Lambda', \pm} =& 
        -i \left({\Delta'}_{\Lambda', \bsymb{k}, -\bsymb{Q}}^{\pdag} \mp \Omega_{j,  \mp \bsymb{Q}}^{\pdag} \right) 
            \ind{\tilde{S}}{j, \bsymb{k}, \bsymb{Q}}{\Lambda, \Lambda', \pm} 
            \nonumber \\
        &-i \sum_{\Lambda''} 
            \left[\ind{g}{j, \bsymb{k}, -\bsymb{Q}}{\Lambda, \Lambda'} \right]^*
            \ind{g}{j, \bsymb{k}, -\bsymb{Q}}{\Lambda'', \Lambda'} 
            N_{j, \bsymb{Q}}^{(\mp)} 
            \tilde{P}_{\Lambda'', \bsymb{k}}^{\pdag}
\end{eqnarray}
%\end{widetext}
with $N_{j, \bsymb{Q}}^{(\mp)} = \left( n_{j, \mp \bsymb{Q}} + \frac{1}{2} \mp \frac{1}{2} \right)$ and ${\Delta'}_{\Lambda', \bsymb{k}, \bsymb{Q}}^{\pdag} = \sum_\bsymb{G} \Delta_{\Lambda', \bsymb{k} + \bsymb{Q} + \bsymb{G}}^{\pdag}$.
Equations~\ref{eq:P1} and \ref{eq:S+-} now form a closed set of equations, that have to be solved for every set of $\bsymb{k}$, $\bsymb{Q}$, $j$, $\Lambda$ and $\Lambda'$.

By formally integrating Eq.~\ref{eq:S+-} we obtain
\begin{eqnarray*}
    \ind{\tilde{S}}{j, \bsymb{k}, \bsymb{Q}}{\Lambda, \Lambda', \pm}(t) 
        &= -i \sum_{\Lambda''} 
        \left[\ind{g}{j, \bsymb{k}, -\bsymb{Q}}{\Lambda, \Lambda'} \right]^*
        \ind{g}{j, \bsymb{k}, -\bsymb{Q}}{\Lambda'', \Lambda'}
        N_{j, \bsymb{Q}}^{(\mp)} \\
    &\times \int\limits_{0}^{t} \dd{\tau} e^{-i\left({\Delta'}_{\Lambda', \bsymb{k}, - \bsymb{Q}} \mp \Omega_{j, \mp \bsymb{Q}} \right) \tau} \tilde{P}_{\Lambda'', \bsymb{k}}(t-\tau)
\end{eqnarray*}
with the initial condition $\ind{\tilde{S}}{j, \bsymb{k}, \bsymb{Q}}{\Lambda, \Lambda', \pm}(0) = 0$, in agreement with the assumption of an ultrashort excitation of the light mode at $t=0$.
The PAPP at time $t$ thus depends on the polarization $\tilde{P}_{\Lambda'', \bsymb{k}}(t-\tau)$ at all earlier times $t-\tau$ with $0 \le \tau \le t$, therefore including memory effects in the dynamics.
Within the 2nd order Born approximation this expectation value can be approximated by $\tilde{P}_{\Lambda'', \bsymb{k}}(t - \tau) \approx \tilde{P}_{\Lambda'', \bsymb{k}}(t) \exp\left[i\Delta_{\Lambda'', \bsymb{k}}\tau \right]$, since all other terms lead to contributions that are higher than 2nd order in the polariton-phonon coupling.
In the theory of open quantum systems this is called the time-convolutionless (TCL) approach \cite{breuer2002the} and it has been found to provide a good description of exciton and polariton spectra in TMDCs  \cite{lengers2020the,lengers2021phon}.

Inserting the previous result into Eq.~\ref{eq:P1} leads to the closed equation for the polarization
\begin{eqnarray}
    \dv{t} \tilde{P}_{\Lambda, \bsymb{k}}(t) = -i \Delta_{\Lambda, \bsymb{k}} \tilde{P}_{\Lambda, \bsymb{k}}(t) - \sum_{\Lambda'} \Gamma_{\Lambda, \Lambda'}(\bsymb{k}, t) \tilde{P}_{\Lambda', \bsymb{k}}(t)
     \label{eq:tcl}
\end{eqnarray}
with time-dependent coefficients
\begin{eqnarray}
    \label{eq:Gamma}
    \Gamma_{\Lambda, \Lambda'}(\bsymb{k}, t) &= \int\limits_{-\infty}^\infty \dd{\Omega} \rho_{\Lambda, \Lambda'}(\bsymb{k}, \Omega) \int\limits_{0}^{t} \dd{\tau} e^{-i\Omega \tau}, 
\end{eqnarray}
where we have introduced the generalized phonon spectral density (gPSD)  $\rho_{\Lambda, \Lambda'}(\bsymb{k}, \Omega)$, defined as 
\begin{eqnarray}\label{eq:rho}
    \rho_{\Lambda, \Lambda'}(\bsymb{k}, \Omega) 
        &= \sum_{{j, \Lambda'', \pm, \bsymb{Q}}} 
            \left[\ind{g}{j, \bsymb{k}, - \bsymb{Q}}{\Lambda, \Lambda''} \right]^*
            \ind{g}{j, \bsymb{k}, -\bsymb{Q}}{\Lambda', \Lambda''} 
        N_{j, \bsymb{Q}}^{(\mp)} \\
    &\times \delta \left[\Omega - \left({\Delta'}_{\Lambda'', \bsymb{k}, - \bsymb{Q}} - {\Delta'}_{\Lambda', \bsymb{k}, \bsymb{0}} \mp \Omega_{j, \mp\bsymb{Q}} \right) \right], \nonumber
\end{eqnarray}
describing phonon-induced transitions within or between the polariton branches at given \moire exciton wave vectors $\bsymb{k}$.
Within this approximation the complete phonon influence is dynamically included in the time-dependent coefficients $\Gamma_{\Lambda, \Lambda'}(\bsymb{k}, t)$ \cite{lengers2020the,lengers2021phon}.

Although memory effects in the polarization have been eliminated in the TCL approach, the coefficients $\Gamma_{\Lambda, \Lambda'}(\bsymb{k}, t)$ still carry information on the time of the excitation process in their time dependence.
This information can be eliminated by additionally performing a Markov approximation, which consists in replacing the coefficients by their long-time limit $\Gamma_{\Lambda, \Lambda'}(\bsymb{k}, t \to \infty)$.
Formally, this is achieved by replacing the time integral in Eq.~\ref{eq:Gamma} with the help of the Dirac identity according to
\begin{eqnarray*}
\lim_{t\to\infty} &\int\limits_{0}^{t} \dd{\tau} e^{-i\Omega \tau}
= \lim_{\eta\to 0} \int\limits_{0}^{\infty} \dd{\tau} e^{-(i\Omega+\eta) \tau} \\
&= \lim_{\eta\to 0} \frac{1}{i\Omega + \eta} = -i \pv \frac{1}{\Omega} +\pi \delta(\Omega).
\end{eqnarray*}
In this limit we obtain $\Gamma_{\Lambda, \Lambda'}^{\rm {Markov}}(\bsymb{k})=\Gamma_{\Lambda, \Lambda'}^{\pdag}(\bsymb{k}, \infty)$ with
\begin{numparts}
\label{eq:Markov}
\begin{eqnarray}
\label{eq:ReMarkov}
    \Re\left[\Gamma_{\Lambda, \Lambda'}^{\rm {Markov}}(\bsymb{k}) \right] &= \pi  \rho_{\Lambda, \Lambda'}^{\pdag}(\bsymb{k}, 0) , \\
\label{eq:ImMarkov}
    \Im\left[\Gamma_{\Lambda, \Lambda'}^{\rm {Markov}}(\bsymb{k})\right] &= - \pv \int\limits_{-\infty}^\infty \dd{\Omega} \frac{\rho_{\Lambda, \Lambda'}^{\pdag}(\bsymb{k}, \Omega)}{\Omega}.
\end{eqnarray}
\end{numparts} 
The real part describes the dephasing due to real (i.e., energy conserving) transitions ($\Omega=0$), while the imaginary part gives rise to energy shifts associated with virtual (i.e., energy non-conserving) transitions.
We can therefore interpret the variable $\Omega$ in the gPSD [Eq.~\ref{eq:rho}] as the energy mismatch of the respective phonon-induced transition.
We will come back to this point later when interpreting the phonon impact on our calculated spectra.
As discussed in \ref{app:sec_pert}, the Markov limit according to Eqs.~\ref{eq:ReMarkov} and \ref{eq:ImMarkov} can be taken as the starting point for a perturbative solution of Eq.~\ref{eq:tcl}.
In Sec.~\ref{sec:comp_pert} we will compare the numerical results with those obtained  from such a perturbative solution, which will provide us with a deeper understanding of the various contributions of the spectrum.

\subsection{Assumptions and parameters}
\label{sec:assumptions}

In \moire crystals the twist angle can be seen as a control parameter that has an impact on the \moire exciton band structure, and therefore the dispersion of the polariton branches \cite{gotting2022moi,fitzgerald2022twi,knorr2022exc,nam2017lat}, the exciton wave functions \cite{gotting2022moi,fitzgerald2022twi}, the phonon dispersion \cite{gao2022sym,koshino2019moi} and the exciton-phonon coupling \cite{koshino2020eff}, i.e., all parameters that contribute to the linear absorption spectrum of Eq.~\ref{eq:spectrum}.

While until now all derivations were general, for the following calculations we consider parameters motivated by \moire excitons in an AA (R-type \cite{rupp2023ima}) stacked MoSe$_2$/WSe$_2$ hetero-bilayer.
For the \moire exciton dispersion we take into account a single homogeneous exciton band with an effective mass of $M = 0.84\, m_0$ \cite{gotting2022moi}, $m_0$ being the free electron mass, describing a quadratic dispersion of the homogeneous excitons.
Note that a constant energy offset will later be chosen such that the \moire exciton at $\bsymb{k} = \bsymb{0}$ is in resonance with the cavity mode at $\bsymb{K} = \bsymb{k} = \bsymb{0}$.
A situation where light-matter coupling is resonant at another wave vector $\bsymb{K} \neq \bsymb{0}$, as for example realized in Ref.~\cite{lundt2016roo} for a TMDC monolayer coupled to an optical cavity, can also lead to interesting effects due to varying proportions of the exciton and light contributions to the polaritons, but will not be discussed in the present work.

The homogeneous excitons experience the periodic \moire potential of  Eq.~\ref{eq:moire_potential}.
We approximate the potential by taking into account the reciprocal lattice vectors  ${\bsymb{G}_j}$ to the first neighbors of the 1st MBZ with $j = 1,\dots,6$.
The numbering is such that ${\bsymb{G}_j}$ and ${\bsymb{G}_{j+1}}$ enclose an angle of $60^\circ$.
The Fourier components of the potential are chosen as $\tilde{V}_{\bsymb{G}_j} = \tilde{V} \exp(i \psi)$ for $j = 1,3,5$, $\tilde{V}_{\bsymb{G}_j} = \tilde{V}\exp(-i \psi)$ for $j = 2,4,6$, and $\tilde{V}_{\bsymb{G}_j} = 0$ otherwise, with $\tilde{V}=11.8\,$meV and phase angle $\psi=79.5^\circ$, \cite{wu2018the,gotting2022moi}.
By solving the eigenvalue equation (\ref{eq:app_eigenvalue_moire}) for every $\bsymb{k}$ in the 1st MBZ numerically, we obtain the \moire exciton dispersion relation $\hbar \omega_{1, \bsymb{k}}$.

\begin{figure}[tb]
    \centering
    \includegraphics[width=0.65\linewidth]{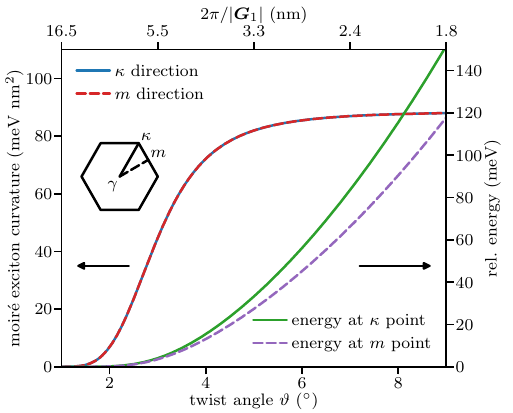}
    \caption{Curvature of the lowest \moire exciton band at $\bsymb{k} = \bsymb{0}$ in the direction of the $\kappa$ (solid blue line) and $m$ point (dashed red line) of the 1st MBZ and exciton energy at the $\kappa$ (solid green line) and $m$ point (dashed violet line) relative to the energy at the $\gamma$ point as a function of the twist angle $\vartheta$ (bottom scale) or, equivalently, the \moire lattice constant $2\pi/|\bsymb{G}_1|$ (top scale). The inset shows a sketch of the 1st MBZ with the definition of the $\kappa$ and $m$ points.}
    \label{fig:fig2}
\end{figure}

Figure~\ref{fig:fig2} shows the curvature of the \moire exciton dispersion $\hbar \omega_{1, \bsymb{k}}$ at the $\gamma$ point in the direction towards the $\kappa$ (blue solid line) and the $m$ (red dashed line) point and the energies at the $\kappa$ (green solid line) and $m$ point (violet dashed line) relative to the $\gamma$ point energy as a function of the twist angle in the range $1^\circ \le \vartheta \le 9^\circ$.
The top axis denotes the absolute value of the real space \moire lattice constant, i.e., the real space \moire periodicity for a given twist angle.
The full exciton dispersion relations for three examplary values of the twist angle $\vartheta$ can be found in Fig.~\ref{app:fig8} in \ref{app:sec_moire-excitons}.

Our first observation is that the curvature is always the same for both depicted high-symmetry directions, showing that the dispersion is isotropic around the $\gamma$ point of the MBZ.
At small twist angles the curvatures at the $\gamma$ point and the energies at the edge of the 1st MBZ vanish, which reflects the flat \moire exciton dispersion in this limit.
When increasing the twist angle, both the curvatures and the energies at the MBZ boundaries increase.
At large twist angles the curvatures saturate, i.e., in this region the \moire exciton dispersion in a certain region around the $\gamma$ point can be approximated in each direction by a parabola with twist-angle independent curvature reflecting the dispersion relation of the homogeneous excitons.
The energies at the $\kappa$ and $m$ points increase with growing angle due to the increasing size of the MBZ.
According to the general properties of band structures, the slope of the exciton dispersion at the MBZ boundary vanishes in the direction perpendicular to the boundary, which leads to an energy maximum at the $\kappa$ point and a saddle point associated with a van Hove singularity in the density of states at the $m$ point.

We set the cavity mode energy $\hbar \omega_\bsymb{0}$ to be in resonance with the \moire exciton band at $\bsymb{k} = \bsymb{0}$.
The \moire exciton-photon coupling constant is set to $\lambda_{1, \bsymb{k}} = 10\,$meV, such that the polariton splitting at $\bsymb{k} = \bsymb{0}$ is on the same order of magnitude as in Refs.~\cite{zhang2021van,fitzgerald2022twi}.

\begin{figure}
    \centering
    \includegraphics[width=0.75\linewidth]{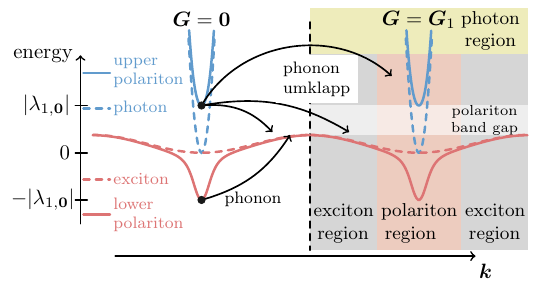}
    \caption{Sketch of exciton (dashed red), photon (dashed blue), and polariton (solid red and blue) dispersion relations in the 1st and 2nd MBZ. The arrows indicate possible phonon-induced transitions within and between polariton branches including phonon umklapp processes. In the right part the exciton, photon, and polariton regions are indicated.}
    \label{fig:fig3}
\end{figure}

A sketch of the exciton (dashed red), photon (dashed blue) and polariton (solid red and blue) dispersion in the 1st and 2nd MBZ is shown in Fig.~\ref{fig:fig3}.
We refer to the region in the MBZ where the exciton-light coupling strongly modifies the dispersion as the polariton region (red area) and to the regions where the dispersion is only weakly affected as exciton (gray area) or photon region (yellow area), respectively.
For clarity these areas are only marked in the 2nd MBZ at $\bsymb{G} = \bsymb{G}_1$ (right).

For the phonons we assume a single acoustic branch with index $j=0$ and linear dispersion $\Omega_{0, \bsymb{Q}} = c_{\rm s} \abs{\bsymb{Q}}$ that is unaffected by the \moire potential and isotropic in all in-plane directions with sound velocity $c_{\rm s}$, i.e., the Debye model.
The phonon wave vector $\bsymb{Q}$ is thus not restricted to the 1st MBZ, but extends over the full Brillouin zone of a single layer.

As derived in \ref{app:sec_phonon_coupling}, the \moire exciton-phonon coupling constant can be written as
\begin{eqnarray}
    \mathcal{G}_{j, \bsymb{k}, \bsymb{Q}}^{(n, n')} &= \left[g_{j, \bsymb{Q}}^{(\rm{e})} F(\mu_{\rm h} \bsymb{Q}) - g_{j, \bsymb{Q}}^{(\rm {h})} F(-\mu_{\rm e}\bsymb{Q}) \right] f_{\bsymb{k}, \bsymb{Q}}^{(n,n')},
    \label{eq:moire_ex-phon}
\end{eqnarray}
where $g_{j, \bsymb{Q}}^{(\rm{e/h})}$ is the coupling constant of electrons/holes, $\mu_{\rm{e/h}} = m_{\rm e/h} / (m_{\rm e} + m_{\rm h})$ is the ratio of the electron (hole) mass $m_{\rm e}$ ($m_{\rm h}$) to the total exciton mass, $F(\bsymb{Q})$ is the form factor of the homogeneous exciton, resulting from the electron-hole attraction (see \ref{app:sec_moire-excitons}), and the form factor $f_{\bsymb{k}, \bsymb{Q}}^{(n,n')}$ originates from the \moire potential confinement.

For the electron/hole-phonon coupling constants in Eq.~\ref{eq:moire_ex-phon} we assume the common deformation potential coupling \cite{lengers2020the}
\begin{eqnarray}
    g_{0, \bsymb{Q}}^{(\rm{e/h})} = \frac{D_{\rm e/h}^{(1)} \abs{\bsymb{Q}}}{\sqrt{2\hbar \rho V \Omega_{0, \bsymb{Q}}}}
\end{eqnarray}
with the mass density of a single layer $\rho$, normalization volume $V$, and deformation potential for electrons/holes in the long-wavelength limit $D_{\rm e/h}^{(1)}$ \cite{jin2014int}, corresponding to the first order perturbation theory of the scattering potential in the phonon wave number.
The phonon parameters of the MoSe$_2$/WSe$_2$ bilayer are approximated by the ones of MoSe$_2$ with density $\rho = 4.26 \times 10^{-7}\,$g/cm$^2$, speed of sound $c_{\rm s} = 4.1\,$nm/ps and average deformation potentials $D_{\rm e}^{(1)} = 2.4\,$eV and $D_{\rm h}^{(1)} = -1.98\,$eV \cite{lengers2020the}.

The exciton form factor $F(\bsymb{Q})$ is obtained from the 1s exciton wave function of the homogeneous exciton.
It satisfies the condition $F(\bsymb{0})=1$ and it is localized roughly on the scale of the inverse of the exciton Bohr radius.
Here, we approximate it by a Gaussian choosing
\begin{eqnarray}
    F(\mu_{\rm e} \bsymb{Q}) \approx F(\mu_{\rm h} \bsymb{Q}) \approx e^{-a^2 \abs{\bsymb{Q}}^2/4}
\end{eqnarray}
with the localization length $a = 2\,$nm.

The \moire form factor $f_{\bsymb{k}, \bsymb{Q}}^{(n,n')}$ in Eq.~\ref{eq:moire_ex-phon} gives rise to an additional twist angle dependence of the \moire exciton-phonon coupling.
It also satisfies the condition $f_{\bsymb{k},\bsymb{0}}=1$.
We assume that the \moire form factor will not strongly affect the general shape of the \moire exciton phonon coupling constant.
Therefore we do not expect new features in the resulting spectra and set $f_{\bsymb{k}, \bsymb{Q}} = 1$ to focus on the twist angle dependence of the \moire exciton dispersion.

The \moire exciton-phonon coupling leads to phonon induced transitions between polariton states either within the same branch or connecting different branches, including both normal and phonon umklapp processes, as sketched schematically in Fig.~\ref{fig:fig3}.

Initially, no excitons and photons are present in the system, while the phonons are in a thermal state with temperature $T$.
We assume that a short laser pulse, applied perpendicular to the cavity surface brings the photons into a coherent state at $\bsymb{k} = \bsymb{0}$ with a small coherent amplitude justifying the limit of small polariton densities.
Since Eq.~\ref{eq:tcl} does not couple different $\bsymb{k}$, the assumption of a perpendicular excitation guarantees that only polariton polarizations with $\bsymb{k}=\bsymb{0}$ are excited in the system at all times.
We additionally introduce a constant damping rate for each polariton polarization by replacing $\hbar \Delta_{\Lambda, \bsymb{k}} \rightarrow \hbar \Delta_{\Lambda, \bsymb{k}} - i\gamma_0$ with $\gamma_0=0.4\,$meV to take other dephasing mechanism like photon loss of the cavity or direct photon emission from the exciton to the continuum of unconfined photon modes into account, which results in non-vanishing spectral line widths.

%------------------------------------------------------------------------
\section{Results}
\label{sec:results}

\subsection{Linear absorption spectra}
\label{sec:real_spec}

We start the results part by discussing linear absorption spectra $\alpha_\bsymb{0}(\omega)$ of \moire exciton polaritons obtained from a full numerical solution of the TCL equation (\ref{eq:tcl}), and analyzing the impact of phonons seen in these spectra as a function of various parameters.
As seen in Eqs.~\ref{eq:tcl}, \ref{eq:Gamma} and \ref{eq:rho}, the spectrum depends on the gPSD $\rho_{\Lambda, \Lambda'}(\bsymb{k}, \Omega)$, which includes the \moire polariton-phonon coupling constants $g_{j, \bsymb{k}, \bsymb{Q}}^{(\Lambda, \Lambda')}$, the phonon dispersion relations $\Omega_{j, \bsymb{Q}}$ and the relative polariton dispersion $\Delta_{\Lambda, \bsymb{k}} = \epsilon_{\Lambda, \bsymb{k}} - \omega_\bsymb{0}$ with respect to the photon mode frequency $\omega_\bsymb{0}$.
Considering all assumptions from the previous section, we notice that $\Delta_{\Lambda, \bsymb{k}}$ is effectively the only quantity that explicitly depends on the twist angle $\vartheta$, as seen in Fig.~\ref{fig:fig2}.

\begin{figure*}[tb]
    \centering
    \includegraphics[width=1.08\linewidth]{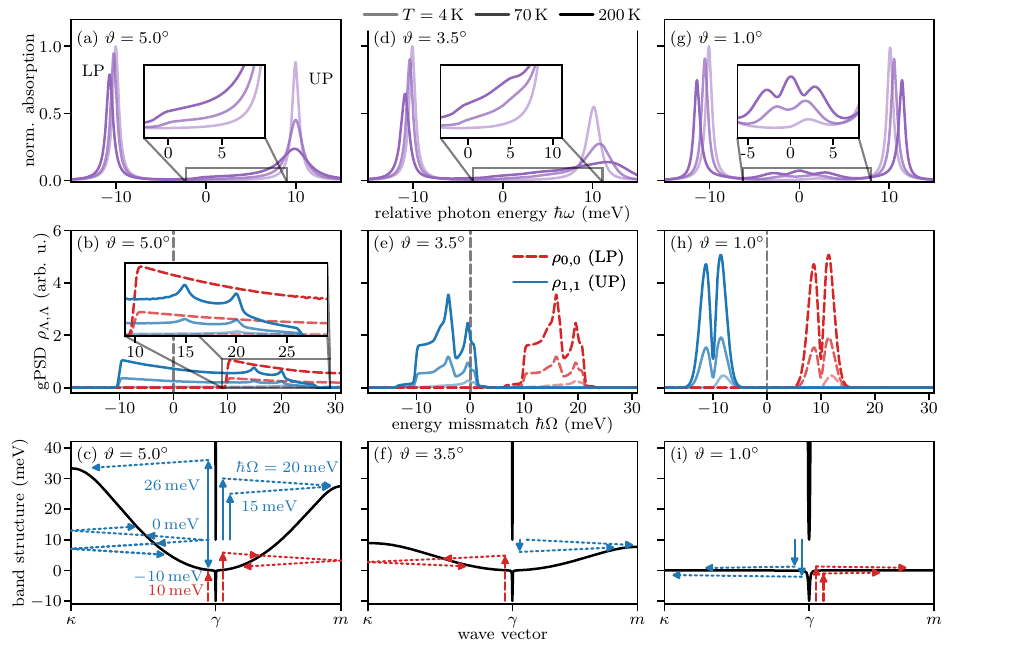}
    \caption{(a, d, g) Linear absorption spectrum, (b, e, h) generalized phonon spectral density (gPSD) at $\bsymb{k}=\bsymb{0}$, and (c, f, i) polariton dispersion at angles (a, b, c) $\vartheta = 5^\circ$, (d, e, f) $\vartheta = 3.5^\circ$ and (g, h, i) $\vartheta = 1^\circ$. The temperatures are $T=4\,$K (light shading), $T=70\,$K (medium shading) and $T=200\,$K (dark shading). The arrows in (c, f, i) sketch possible real and virtual phonon transitions including umklapp processes. In the middle and bottom row blue lines and arrows refer to the upper polariton branch, red ones to the lower polariton branch.}
    \label{fig:fig4}
\end{figure*}

In Fig.~\ref{fig:fig4}, the top row (a, d, g)  shows the linear absorption spectra for the temperatures $T=4, 70, 200\,$K (from bright to dark colors) and twist angles $\vartheta = 5^\circ, 3.5^\circ$, and $1^\circ$ (from left to right).
The middle row (b, e, h) depicts the corresponding gPSDs $\rho_{\Lambda, \Lambda}(\bsymb{k}, \Omega)$ at $\bsymb{k}=\bsymb{0}$ for the lower ($\Lambda=0$, dashed red) and the upper ($\Lambda=1$, solid blue) polariton as a function of the energy mismatch $\hbar \Omega$.
Note that for our set of parameters, in particular due to the resonant exciton-photon coupling, $\rho_{0, \Lambda}(\bsymb{k}, \Omega)$ and $\rho_{1, \Lambda}(\bsymb{k}, \Omega)$ are real and identical (not shown), therefore it is sufficient to show and discuss the diagonal elements $\rho_{\Lambda, \Lambda}$.
Finally, the bottom row (c, f, i) illustrates the polariton dispersion relation of the respective structure with exemplary phonon transitions (arrows).
The dispersion relations immediately reveal the qualitative differences between the three chosen angles: While for $\vartheta = 5^\circ$ there is a large overlap in the energies of the lower (LP) and upper (UP) polariton branches, for $\vartheta = 1^\circ$ the LP dispersion is essentially flat in the exciton region, resulting in a pronounced energy gap between LP und UP.
The intermediate angle $\vartheta = 3.5^\circ$ marks the transition between these two limiting cases, where the top of the LP branch is approximately at the same energy as the bottom of the UP branch.

We start with the discussion of the spectra at $\vartheta = 5^\circ$ [Fig.~\ref{fig:fig4}~(a)].
Here, in the spectra we find two main peaks, directly reflecting absorption by the UP (right peak) and the LP (left peak).
The peak position of the LP slightly depends on temperature, while its width is mainly unaffected.
In contrast, the peak position of the UP remains essentially constant, but it gets significantly broader with increasing temperature.
Furthermore, we find a continuous background extending from the upper peak to lower energies, highlighted in the inset.
All these effects are induced by phonon-assisted intra- and inter-band transitions and can be explained using the gPSD in Fig.~\ref{fig:fig4}~(b) in combination with the polariton dispersion in (c).

The gPSD for the UP, $\rho_{11}$ [solid blue line in Fig.~\ref{fig:fig4}~(b)], is a broad distribution that abruptly starts at $\hbar \Omega \approx -10\,$meV, slowly decays until a sharp cut-off at $\hbar \Omega \approx 26\,$meV, and exhibits two distinct peaks at $\hbar \Omega \approx 15\,$meV and $20\,$meV.
The left peak vanishes at low temperatures, which will be discussed later.
The gPSD for the LP, $\rho_{00}$ (dashed red line), has the same shape as the one of the UP but it is shifted by approximately $20\,\rm{meV} = 2 \lambda_{1,\bsymb{k}}$ to higher energies (not shown entirely).
At first glance, in comparison to the spectra in (a), the energetic ordering of the gPSD contributions seems counterintuitive.
However, it can be understood by having a look at the phonon processes entering the gPSD.

As can be seen from the delta-distribution in its definition in Eq.~\ref{eq:rho}, $\rho_{\Lambda,\Lambda'}(\bsymb{k},\Omega)$ describes phonon-induced transitions from an initial polariton state with energy $\hbar \epsilon_{\Lambda', \bsymb{k}}$ to all possible final states with energies $\hbar \epsilon_{\Lambda'', \bsymb{k}-\bsymb{Q}-\bsymb{G}}$ under the emission or absorption of a phonon with energy $\hbar \Omega_{j,\pm\bsymb{Q}}$ and $\Omega$ describes the deviation from energy conservation, i.e., the energy mismatch of the process, in agreement with the finding in Eq.~\ref{eq:ReMarkov} that $\Omega=0$ corresponds to real, i.e., energy-conserving transitions, which survive in the Markov limit.
Based on this interpretation, exemplary phonon-induced intra- and inter-band transitions starting from $\bsymb{k}=\bsymb{0}$ are sketched in Fig.~\ref{fig:fig4}~(c).
The phonon emission or absorption is sketched by dotted arrows with a slope determined by the phonon dispersion relation, i.e., by the sound velocity.
Dotted arrows with a downward slope represent phonon emission processes and those with upward slope phonon absorption.
Arrows that are 'reflected' at the boundary of the MBZ are umklapp processes which will have an impact on the gPSD in particular for small Brillouin zones, where the involved phonon wave vector $\bsymb{Q}$ exceeds the 1st MBZ.
The energy mismatch $\hbar \Omega$ is indicated by a vertical, solid or dashed arrow.
A vanishing energy mismatch, i.e., $\hbar \Omega = 0$, corresponds to energy-conserving phonon transitions, while $\hbar \Omega \neq 0$ describes off-resonant transitions which are in general suppressed due to the $1/\Omega$ scaling in Eq.~\ref{eq:Gamma}.
The blue (red) arrows correspond to transitions starting at the $\gamma$ point of the UP (LP), where the small horizontal offset of the arrow is chosen for a better visibility.

With this in mind we can now explain the general shape of the gPSD in Fig.~\ref{fig:fig4}~(b),
starting with the gPSD for the UP (solid blue lines).
The abrupt onset at $\hbar \Omega \approx -10\,$meV and end at $\hbar \Omega \approx 26\,$meV can be traced back to the finite energy width of the LP band.
The lower cut-off directly corresponds to transitions starting from the $\gamma$ point of the UP and ending in the exciton region at the energy $0\,$meV of the LP.
The high energy cut-off stems from processes which end at the edge of the 1st MBZ ($\kappa$ point) in the LP branch.
The energy spread of the \moire exciton dispersion thus determines the width of the gPSD.
Strictly speaking, transitions to the polariton region of the LP are also possible, and they are in fact present in the region between $-20\,$meV and  $-10\,$meV.
However, due to the large curvature of the LP branch and the resulting very small density of states in the polariton region these contributions have a negligible weight in the gPSD.
Consequently, the general shape of the gPSD is dominated by transitions to the exciton part.

Energy-conserving transitions, i.e., transitions with $\Omega=0$, are possible from the bottom of the UP to the LP branch, both by normal and by umklapp processes, as indicated by the dotted lines on the left side, starting from the bottom of the UP branch.
The non-vanishing value of the gPSD of the UP at $\hbar \Omega = 0$ gives rise to a strong phonon impact on the UP resonance in the spectrum [Fig.~\ref{fig:fig4}~(a)], resulting in its pronounced broadening with increasing temperature.

We now come back to the peaks in the gPSD seen in Fig.~\ref{fig:fig4}~(b) at $\hbar \Omega \approx 15\,$meV and $20\,$meV.
They describe transitions to the flat part of the LP branch close to the edge of the 1st MBZ, namely to the $m$ point of the LP, while absorbing or emitting a phonon.
This leads to van Hove singularities in the gPSD, but at this point the processes are strongly energy non-conserving and therefore they do not have a significant impact on the spectrum in Fig.~\ref{fig:fig4}~(a).
From the asymmetry in the relative peak heights in (b) for small temperatures we can conclude, that the higher energetic peak, i.e., the one  characterized by $\hbar \Omega = 20\,$meV in (c), mainly belongs to phonon emission and the lower one labeled by $\hbar \Omega = 15\,$meV to phonon absorption processes.
In the limit $T \rightarrow 0\,$K phonon absorption processes are not possible anymore, while phonon creation remains.

When comparing the gPSD of the UP, $\rho_{11}$ (solid blue line), with the one of the LP, $\rho_{00}$ (dashed red line), we directly see that they mainly differ by their energetic positions.
Because transitions to the polariton part are again negligible due to the small density of states, the gPSD essentially starts at an energy of $10\,$meV with transitions to the bottom of the exciton part of the dispersion and it ends at an energy of $46\,$meV with transitions to the $\kappa$ point (not shown).
Similar to the case of the UP gPSD, peaks reflecting the van Hove singularities appear, now at energies $\hbar \Omega \approx 35\,$meV and $40\,$meV and therefore outside the plotted energy range.

While the gPSD of the UP is non-vanishing around $\hbar \Omega = 0$, i.e., energy-conserving phonon transitions are possible, the gPSD of the LP is shifted to higher energies, such that only energy non-conserving transitions appear in the spectrum.
These energy non-conserving transitions give rise to the energy renormalization [see Eq.~\ref{eq:ImMarkov}], i.e., to the temperature-dependent shift of the LP peak in the spectrum in (a).
This explains why the LP peak exhibits mainly an energy shift while the UP peak shows a prominent broadening.
Energy non-conserving transitions also lead to the continuous background at energies below the UP peak in the spectrum, representing a phonon sideband.
Its origin and its specific shape will be discussed later.

The spectra for a twist angle of $\vartheta = 3.5^\circ$ are shown in Fig.~\ref{fig:fig4}~(d).
Overall, we find the same features as before, but the UP peak is broader than in the spectra for $\vartheta = 5^\circ$ and some additional structures emerge in the phonon sideband region, as highlighted in the inset.
When investigating the corresponding gPSD in (e) we again find two pronounced peaks for the UP (solid blue lines) and LP (dashed red lines), reflecting transitions to the van Hove singularities at the $m$ points, which are now shifted to lower energies due to the reduction of the size of the MBZ and the decreased curvature of the \moire exciton dispersion.
This can also be seen in the polariton dispersion in Fig.~\ref{fig:fig4}~(f), where the energy of the LP at the edge of the MBZ is much smaller compared to the band structure for $\vartheta = 5^\circ$ and it is now close to the energy of the UP at the $\gamma$ point.
Most of the gPSD peaks are energy non-conserving, such that they have a relatively small impact on the spectra.
They lead to the small structures in the continuous phonon sideband below the UP peak and contribute to the shift of the polariton peak positions.
However, the peak in the gPSD of the UP at $\hbar \Omega \approx 0$ is energy-conserving and therefore leads to the strong broadening of the UP peak in the spectra in Fig.~\ref{fig:fig4}~(d).
Note that this peak in the gPSD emerges from a phonon emission process [highest dashed blue arrow in (f)], explaining the significant spectral broadening of the UP peak even at low temperatures.

For almost entirely flat bands at $\vartheta = 1^\circ$ the spectra are shown in Fig.~\ref{fig:fig4}~(g).
We find that the UP and LP peaks are not significantly broadened and the continuous phonon background is replaced by one peak for $T=4\,$K, two for $T=70\,$K, and three for $T=200\,$K in the middle between the polariton peaks.
In the gPSD in (h) we do not find energy-conserving transitions anymore, i.e., we have vanishing values at $\hbar \Omega = 0$, only a double peak structure centered at approximately $-10\,$meV for the UP (blue) and around $+10\,$meV for the LP (red).
The transitions leading to these peaks are sketched in the corresponding polariton dispersion in (i).
Due to the opening of a polariton band gap, energy-conserving transitions to the exciton part are not possible anymore, neither for the LP nor for the UP.
This explains the small width of the UP peak in (g) compared to the previous twist angles and the dominance of the energy shifts for both peaks.

The peak structure in the phonon sideband region of the spectrum between the two polariton peaks can also be traced back to the gPSD.
It can be understood best on the basis of the perturbative solution described in the following section.

\subsection{Comparison with the Markovian limit and a perturbative solution}
\label{sec:comp_pert}
In the previous section the spectra have been obtained from a numerical solution of Eq.~\ref{eq:tcl}.
There, we have seen that the phonon impact on the \moire exciton-polaritons depends non-trivially on the twist angle.
On the basis of this solution, together with the corresponding gPSD, we were able to clearly interpret important features like the energy shifts and the broadenings of the polariton peaks.
However, other features, in particular the detailed structure of the phonon sidebands turned out to be not so obvious.

The basic equation (\ref{eq:tcl}) of our model is obtained in the TCL scheme; it is time-local but it is still non-Markovian.
In Sec.~\ref{sec:eom} we have derived the Markovian limit and we have seen that -- as usual -- this limit gives rise to energy shifts and broadenings of the spectral lines.
In the following we will use the Markovian limit as the starting point for a perturbative solution of Eq.~\ref{eq:tcl}.
We will then compare the exact numerical results with various orders of the perturbative solution, allowing us to obtain a better understanding of the role of non-Markovian effects and, in particular, of the sideband structure in the spectra.

Starting point for the perturbative solution is the separation of the time-dependent coefficients $\Gamma_{\Lambda, \Lambda'}(\bsymb{k}, t)$ from Eq.~\ref{eq:Gamma} into a time-independent Markovian part given by Eq.~\ref{eq:Markov}
and a time-dependent non-Markovian correction $\delta\Gamma_{\Lambda, \Lambda'}(\bsymb{k}, t)$ according to 
\begin{eqnarray}
\delta\Gamma_{\Lambda, \Lambda'}(\bsymb{k}, t) = \Gamma_{\Lambda, \Lambda'}(\bsymb{k}, t) - \Gamma^{\rm{Markov}}_{\Lambda, \Lambda'}(\bsymb{k}).
\label{eq:separation}
\end{eqnarray} 
This leads to the equation of motion
\begin{eqnarray}
    \dv{t} \tilde{P}_{\Lambda, \bsymb{k}}(t) &= \sum_{\Lambda'} \left[ -i A_{\Lambda, \Lambda'}(\bsymb{k}) - \delta \Gamma_{\Lambda, \Lambda'}(\bsymb{k}, t) \right] \tilde{P}_{\Lambda', \bsymb{k}}(t),
    \label{eq:pol_split}
\end{eqnarray}
where
\begin{eqnarray}
    A_{\Lambda, \Lambda'}(\bsymb{k}) &= \Delta_{\Lambda, \bsymb{k}} \delta_{\Lambda, \Lambda'} - i \Gamma_{\Lambda, \Lambda'}^{\rm{Markov}}(\bsymb{k}) 
\label{eq:AMarkov}
\end{eqnarray} 
includes the bare polariton part and the Markovian part of the phonon interaction.
Neglecting the non-Markovian part, Eq.~\ref{eq:pol_split} is a linear differential equation with time-independent coefficients in an $(N+1)-$dimensional space, which for sufficiently small values of $N$ can be solved analytically by a suitable rotation in that space, leading to Lorentzian peaks in the rotated basis.
Thus we can deduce, that the Markovian part $\Gamma_{\Lambda, \Lambda'}^{\rm{Markov}}$ leads to a polaronic energy shift due to its imaginary part, a dephasing due to its real part, and a mixing of the polariton branches due to its off-diagonal nature.
The non-Markovian part $\delta \Gamma_{\Lambda, \Lambda}(\bsymb{k}, t)$, which will be treated as a perturbation, then dynamically includes corrections to the Markovian line shifts and line broadenings and it gives rise to phonon sidebands \cite{lengers2020the}.
In the following we will give a brief overview of the main steps in the derivation of this perturbative approach.
Details are given in \ref{app:sec_pert}.

First, a change of basis is performed, which diagonalizes the Markovian part and allows us to obtain an analytical solution of this part.
This analytical solution corresponds to the 0th order in the perturbation.
Afterwards we perform a Laplace transform [see \ref{app:Laplace}], which results in a Fredholm equation (\ref{app:fredholm}) \cite{courant1953met}, where the non-Markovian part contributes via a convolution operator $\mathcal{K}_{\bsymb{k}}$ [see \ref{app:pert_operator}].
By treating the non-Markovian part as a small perturbation we obtain an iterative solution of the Fredholm equation [see Eqs.~\ref{app:pert_iteration0} and~\ref{app:pert_iterationn}].
For more details we refer to  \ref{app:sec_pert}.

As already mentioned, in the 0th order, i.e., in the Markovian limit, the exciton-phonon coupling leads to polaron-shifted polariton peaks broadened by the dephasing due to energy-conserving, phonon-induced transitions.
The energy renormalization mainly depends on the area under the function $\rho_{\Lambda, \Lambda'}(\bsymb{0}, \Omega) / \Omega$ while the dephasing is determined by $\rho_{\Lambda, \Lambda'}(\bsymb{0}, 0)$.
The non-Markovian part then enters in the $n$th order with $n\geq1$ and involves a convolution of $\rho_{\Lambda, \Lambda'}(\bsymb{0}, \Omega) / \Omega$ with the solution in the $(n-1)$th order and a correction term involving $\rho_{\Lambda, \Lambda'}(\bsymb{0}, 0)$.
\begin{figure*}[t]
    \centering
    \includegraphics[width=1.0\linewidth]{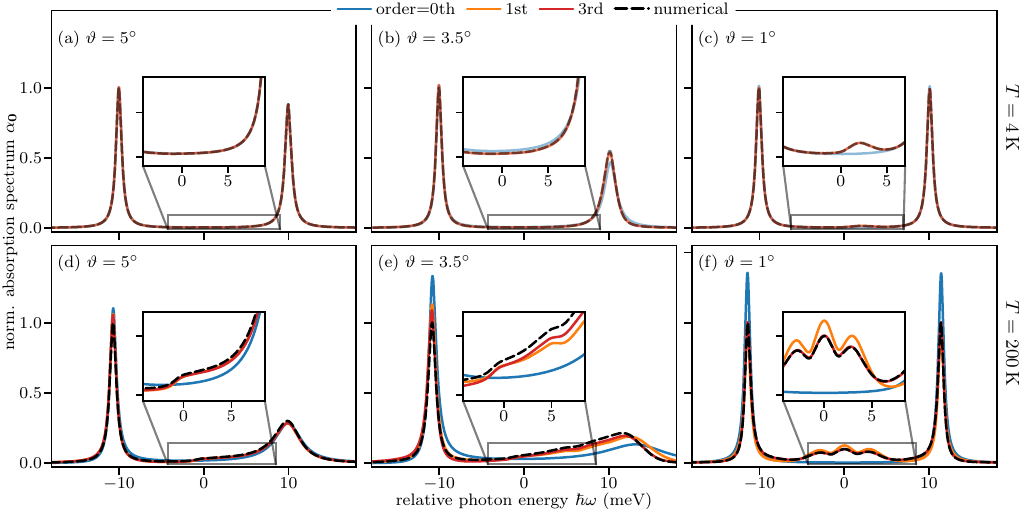}
    \caption{Linear absorption spectra obtained from the perturbative solution (solid lines) in 0th (blue), 1st (orange) and 3rd order (red), compared to the numerically obtained ones (dashed black lines) at temperatures $T=4\,$K (a, b, c) and $T = 200\,$K (d, e, f) for twist angles $5^\circ$ (a, d), $3.5^\circ$ (b, e), and $1^\circ$ (c, f).}
    \label{fig:fig5}
\end{figure*}

In Fig.~\ref{fig:fig5} we compare the spectra obtained from the numerical solution of Eq.~\ref{eq:tcl} (dashed black lines) with the ones obtained by the perturbative solution according to \ref{app:fredholm} in increasing perturbation orders (solid blue, orange and red lines).
The temperatures $T=4\,$K (top row) and $T=200\,$K (bottom row) and twist angles of $\vartheta = 5^\circ$ (a, d), $\vartheta = 3.5^\circ$ (b, e), and  $\vartheta = 1^\circ$ (c, f)  are the same as in Fig.~\ref{fig:fig4}.

We find that at low temperature ($T=4\,$K, top row) both polariton peaks are already very well reproduced by the 0th order, i.e., by the Markovian limit (blue solid line) for all twist angles.
In particular, for a large angle of  $\vartheta = 5^\circ$ [Fig.~\ref{fig:fig5}~(a)] all curves are essentially indistinguishable showing that non-Markovian effects are completely negligible here.
For $\vartheta = 3.5^\circ$ [Fig.~\ref{fig:fig5}~(b)] the Markovian result slightly underestimates the height of the UP peak and overestimates the spectrum between the two peaks, but already the 1st order corrections lead to a perfect agreement with the numerical result.
Higher order corrections turn out to be negligible here.
For $\vartheta = 1^\circ$ [Fig.~\ref{fig:fig5}~(c)] the peaks are again well reproduced in the Markovian limit, but the full solution exhibits a weak additional peak between the two polariton peaks.
This is a phonon sideband and it reflects the absorption of a photon under the simultaneous emission of a phonon.
It can be understood from the corresponding gPSD, shown in Fig.~\ref{fig:fig4}~(h) (light curves).
We see that the gPSD of the UP (blue) exhibits a peak at $\hbar \Omega \approx -8\,$meV and the gPSD of the LP (red) a peak at $\hbar \Omega \approx 12\,$meV.
Since the 1st order correction is given by a convolution of the gPSD with the 0th order spectrum, these peaks lead to a phonon sideband of the UP about $8\,$meV below the main UP peak and a phonon sideband of the LP about $12\,$meV above the main LP peak.
Since the main peaks are separated by $20\,$meV, the two sidebands overlap resulting in the single additional peak in Fig.~\ref{fig:fig5}~(c) at $\hbar \omega \approx 2\,$meV.
Again, the 1st order correction already leads to a perfect agreement with the numerical solution, showing that higher order corrections are negligible.

The situation changes at the higher temperature of $T=200\,$K (bottom row), where we find clear deviations of the Markovian result (blue lines) from the results of the full numerical solution (black dashed lines).
At a large twist angle of $\vartheta = 5^\circ$ [Fig.~\ref{fig:fig5}~(d)] the UP and the LP peak, in particular their asymmetry, are well reproduced already in the Markovian limit.
Between these peaks a weak but clearly visible broad phonon sideband appears, which reflects the broad gPSD in Fig.~\ref{fig:fig4}~(b) (dark curves) with the rather sharp onset at $-10\,$meV for the UP and $10\,$meV for the LP.
Again, the sidebands of LP and UP overlap; they are characterized by an onset at $\hbar \omega \approx 0$ and they are completely described by the 1st order correction with vanishing higher order contributions.

At an intermediate twist angle of $\vartheta = 3.5^\circ$ [Fig.~\ref{fig:fig5}~(e)] we find that the Markovian approximation strongly overestimates the broadening of the UP peak.
In fact,  the gPSD in Fig.~\ref{fig:fig4}~(e) (dark curves) is characterized by a rather large value of $\rho_{1,1}(\bsymb{0},0)$ leading to a strong dephasing.
This is reduced by non-Markovian effects, which leads to an increase of the spectrum at the UP peak and a decrease at the LP peak.
Furthermore, a sideband with an onset at $\hbar \omega \approx 0$ and a peak at  $\hbar \omega \approx 6\,$meV reflecting the shape of the gPSD shows up.
In this case of a quite efficient exciton-phonon coupling, even the 3rd perturbation order is not sufficient to perfectly reproduce the numerical solution.

Finally, at a small twist angle of $\vartheta = 1^\circ$ [Fig.~\ref{fig:fig5}~(f)] the positions and the widths of the main peaks are well reproduced by the 0th order, i.e., the Markovian approximation.
However, we find a redistribution of oscillator strength to the sideband region leading to a reduced height of the main peaks.
The sideband now exhibits a three peak structure, which is qualitatively well described in 1st order and quantitatively reproduced in 3rd order perturbation theory.
At first sight this seems surprising since the gPSD in Fig.~\ref{fig:fig4}~(h) (dark curves) exhibits a two-peak structure both for the UP and the LP, centered around $\mp 10\,$meV.
However, we notice that the main peaks in the spectrum now exhibit pronounced shifts by about $\pm 1.5\,$meV for the UP and LP, respectively.
Therefore, the sidebands corresponding to the LP and the UP do not completely overlap anymore completely.
Instead, the higher energy peak of the LP sideband overlaps with the lower energy peak of the UP, resulting in the three-peak structure with an enhanced middle peak.

\subsection{Twist angle dependence}
\label{sec:twist_dep}

In the previous sections we have studied the phonon influence on the optical absorption spectra of \moire exciton polaritons for several twist angles and temperatures.
We have seen that the phonon coupling essentially modifies the spectra with respect to three key features: (i) it leads to a temperature dependent broadening of the polariton peaks; (ii) it gives rise to temperature dependent shifts of the peaks; and (iii) it leads to the appearance of phonon sidebands in the spectra, which manifest themselves either in new peaks or shoulders in the region between the main polariton peaks.
We were able to clearly interpret these features on the basis of our TCL approach which revealed that the broadening and the shift of the peaks are in most cases well described by the Markovian limit, while the phonon sidebands as well as corrections to the peak heights and widths are consequences of the non-Markovian contributions to the equations of motion.
In this section we take a more general view on the twist angle dependence of the phonon influence.
First, we analyze the deviations between the perturbative and the numerical solution and then we examine the twist angle dependence of width and position of the LP and UP peaks in the spectrum.

\begin{figure}[t]
    \centering
    \includegraphics[width=0.65\linewidth]{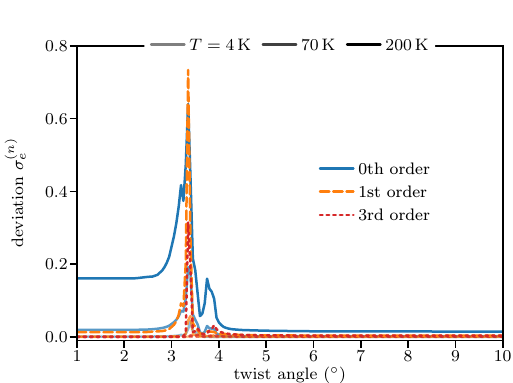}
    \caption{Normalized deviation of the perturbative method from the numerical solution (see Eq.~\ref{eq:deviation}) as a function of the twist angle $\vartheta$ for several temperatures and perturbation orders.}
    \label{fig:fig6}
\end{figure}

In Fig.~\ref{fig:fig6} we plot the normalized deviation of the $n$th perturbation order compared to the numerical solution, calculated by
\begin{eqnarray}
    \sigma_{\rm e}^{(n)} = \frac{\int \dd{\omega} \abs{\alpha_\bsymb{0}^{(n)}(\omega) - \alpha_\bsymb{0}^{(\rm{num.})}(\omega)}^2}{\int \dd{\omega} \abs{\alpha_\bsymb{0}^{(\rm{num.})}(\omega)}^2}
    \label{eq:deviation}
\end{eqnarray}
as a function of the twist angle for several temperatures and perturbation orders.
We can clearly distinguish three regimes of twist angles: In the regimes of large ($\vartheta \gtrsim 4.5^\circ$) or small ($\vartheta \lesssim 2.5^\circ$) twist angles the deviation of a given order $n$ is essentially independent of the twist angle while in the intermediate regime  ($2.5^\circ \lesssim \vartheta \lesssim 4.5^\circ$) pronounced variations as a function of the twist angle appear.

In the regime of large twist angles ($\vartheta \gtrsim 4.5^\circ$) already the 0th order (solid blue), i.e., the Markovian limit, is in general a good approximation.
Only for the highest temperature there are some deviations in the 0th order, which are essentially removed by the 1st order (dashed orange).
The reason is that in this regime the exciton band width is larger than the polariton band gap, such that there are always energy-conserving transition between the UP and the LP branch.
These reflect the main phonon influence and are typically well described in the Markovian limit.
Only at higher temperatures a phonon sideband builds up, which is then well described in the 1st order.

At small twist angles ($\vartheta \lesssim 2.5^\circ$) the deviations of the Markovian limit from the full result are larger, in particular at higher temperatures, which can be traced back to pronounced phonon sidebands in this regime.
However, the perturbative approach converges rather quickly and in 3rd order (dotted red) the deviations from the numerically obtained result become negligibly small for all temperatures.

In the intermediate range of twist angles ($2.5^\circ \lesssim \vartheta \lesssim 4.5^\circ)$ pronounced deviations of the Markovian limit occur and also the convergence of the perturbative solution is less good.
In particular, at higher temperatures even the 3rd order still exhibits deviations of up to almost 30\%.
In this regime the impact of phonons on the spectrum is particularly strong because now energy-conserving transitions to the van Hove singularities are relevant.
Due to these efficient scattering channels also the non-Markovian terms are rather large resulting in the poor convergence of the perturbative approach.

\begin{figure}[tb]
    \centering
    \includegraphics[width=0.65\linewidth]{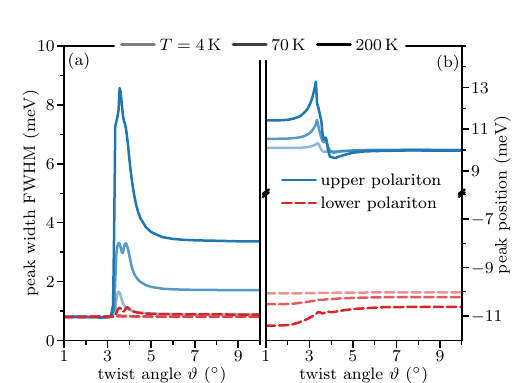}
    \caption{Twist angle scan of the width (a) and position (b) of the UP and LP peaks for different temperatures.}
    \label{fig:fig7}
\end{figure}

Finally, we analyze the phonon impact on the width and position of the UP and LP peaks.
Without coupling to phonons, these are two Lorentzian peaks at $\pm 10\,$meV with a width of about $0.8\,$meV, determined by the background dephasing $2 \hbar \gamma_0$.
Figure~\ref{fig:fig7} shows the phonon impact on the peak width in (a) and the peak position in (b) of the UP (blue) and LP (red) peak as a function of the twist angle, again for different temperatures (with bright colors for low and dark colors for high temperatures).

We can clearly distinguish the same three regimes of twist angles as in Fig.~\ref{fig:fig6}: In the regimes of large ($\vartheta \gtrsim 4.5^\circ$) or small ($\vartheta \lesssim 2.5^\circ$) twist angles the peak widths and peak positions are essentially independent of the twist angle while in the intermediate regime  ($2.5^\circ \lesssim \vartheta \lesssim 4.5^\circ$) pronounced variations of both quantities as functions of the twist angle are seen.

For small twist angles ($\vartheta \lesssim 2.5^\circ$) we find a negligible phonon influence on the peak width.
The peak width is independent of both twist angle and of temperature; it is essentially given by the constant background dephasing rate $2 \hbar \gamma_0$.
Due to the flat exciton band and the resulting energy gap in the polariton dispersion relation no real, i.e., energy-conserving, phonon-induced transitions are possible.
In contrast, the peak positions exhibit pronounced, temperature dependent shifts leading to an increasing separation between LP and UP peak with increasing temperature.
This can be understood from the gPSD in Fig.~\ref{fig:fig4}~(h).
The UP and the LP contribution are almost mirror symmetric, the UP part being located entirely at negative energies and the LP part at positive energies.
Therefore, the Markovian energy shifts according to Eq.~\ref{eq:ImMarkov} are of similar size but with opposite sign, as is indeed seen in Fig.~\ref{fig:fig7}~(b).

In contrast, for large twist angles ($\vartheta \gtrsim 4.5^\circ$) we find a strongly temperature dependent width of the UP peak (a) and energy shift of the LP peak (b), while the impact on the width of the LP and the shift of the UP peak is negligible.
Here, energy-conserving transitions from the bottom of the UP branch to the LP branch are always possible while from the bottom LP branch no such transitions to regions with large density of states are possible.
On the other hand, as can be seen in Fig.~\ref{fig:fig4}~(b), the gPSD of the LP is non-vanishing only for positive energies leading to the observed energy shift, while the gPSD of the UP is almost symmetric around $\Omega=0$, which leads to efficient cancellations in Eq.~\ref{eq:ImMarkov}.

At intermediate twist angles  ($2.5^\circ \lesssim \vartheta \lesssim 4.5^\circ$) we find a significantly increased energy shift and broadening of the UP peak, while the LP peak is only weakly affected.
This is caused by the pronounced impact of phonon transitions to the boundary of the 1st MBZ, giving rise to van Hove singularities at or close to $\Omega=0$ in the gPSD.
At the same time, the gPSD is still very asymmetric with respect to $\Omega=0$, leading to a strong energy shift.
In the recently introduced language of twisted 2D materials one might therefore call this regime the \textit{magic angle for phonon scattering}, which becomes particularly prominent here.
In \ref{app:twist} we present further quantities, namely the gPSD at $\Omega=0$ and the phonon sideband strength, that also show the transition from small to large twist angles around $\vartheta = 3.5^\circ$.

\section{Conclusion}
\label{sec:conclus}

In this paper we have derived a numerical model for the linear absorption spectrum of excitons in a TMDC \moire lattice strongly coupled to an optical cavity, where we found that the impact of phonons depends on several quantities, e.g., the \moire exciton dispersion, \moire exciton wave functions and the exciton-phonon coupling.
These quantities can directly or indirectly be adjusted via the twist angle of the heterostructure.
In the present paper we focused on the phonon impact associated with the varying \moire exciton dispersion.
The phonon influence can be expressed in terms of a generalized phonon spectral density, which turned out to be a useful concept for the understanding and interpretation of the spectra.
In a comparison with a perturbative solution we were able to identify Markovian processes as the main origin of the energy renormalizations of the polariton peaks and their broadenings.
Non-Markovian processes lead to corrections of the Markovian shifts and broadenings.
Most importantly, however, they give rise to phonon sidebands in the spectral region between the polariton peaks.

We found fundamentally different optical spectra for small, intermediate and large twist angles with a strongly increased phonon impact for intermediate values around $3.5^\circ$ which one might call \textit{magic angle for phonon scattering}.
For different designs a different value for the \textit{magic angle} is possible.
On the one hand, spectra at small twist angles are governed by the presence of a polariton band gap, suppressing phonon transitions and thus leading to sharp polariton peaks.
Between the UP and the LP peak characteristic peaked phonon sidebands appear, reflecting phonon assisted transitions to the flat, excitonic part of the polariton dispersion.
On the other hand, spectra for large twist angles are dominated by energy-conserving phonon transitions leading to a broadening of the UP peak.
Here, the phonon sidebands are broad structures reflecting the wide range of possible phonon transitions.
Around the \textit{magic angle for phonon scattering} phonon processes are most pronounced.
This is caused by energy-conserving transitions to the edge of the 1st MBZ, i.e., into van Hove singularities, which then lead to increased phonon scattering resulting in a pronounced peak broadening of the UP.
Due to the efficient coupling to phonons the convergence of the perturbative approach is reduced.

Our results show that the rich interplay between \moire exciton-polaritons, which for small twist angles might be considered as qubit arrays for quantum technological applications \cite{yu2017moi}, and phonons is a key ingredient to fully understand the detected optical response.
Using the twist angle in the heterostructure as a control parameter the phonon impact might be promoted or suppressed, depending on the desired functionality of the pursued device.

\section*{Data availability statement}
    The data that support the findings of this study are available upon reasonable request from the authors.
    The shown data is not measured. Everything is simulated and can be reproduced by the information given in the text.

\ack
Financial support of the Alexander von Humboldt foundation in the framework of a research group linkage grant with the group of Pawe{\l} Machnikowski (Wroc{\l}aw University of Science and Technology, Poland) is gratefully acknowledged.

\appendix

\section{Derivation of \moire excitons}
\label{app:sec_moire-excitons}
In this section we derive the Hamiltonian for excitons in a \moire potential starting from the electron-hole picture for a homogeneous 2D semiconductor.
This derivation basically follows the supplementary material of Ref.~\cite{brem2020tun,knorr2022exc}.
The Hamiltonian of electrons/holes in a \moire potential
\begin{eqnarray}
    V_{\rm{m}}^{\rm{e/h}}(\bsymb{r}) = \sum_{j} V_{\bsymb{G}_j}^{(\rm{m, e/h})} e^{i\bsymb{G}_j \cdot \bsymb{r}},
    \label{eq:app_moire_potential}
\end{eqnarray}
where $\bsymb{G}_j$ is the $j$th reciprocal \moire lattice vector and $V_{\bsymb{G}_j}^{(\rm{m, e/h})}$ is the corresponding Fourier component of the \moire potential for electrons/holes, is given by
\begin{eqnarray*}
    H &= \sum_\bsymb{K} \left[\epsilon_\bsymb{K}^{(\rm{e})} c_\bsymb{K}^\dagger c_\bsymb{K}^{\pdag} + \epsilon_\bsymb{K}^{(\rm{h})} d_\bsymb{K}^\dagger d_\bsymb{K}^{\pdag} \right] \\
        &\mbox{\quad} - \sum_{\bsymb{K}, \bsymb{K}', \bsymb{Q}} V_\bsymb{Q} c_\bsymb{K}^\dagger d_{\bsymb{K}'}^\dagger d_{\bsymb{K}' - \bsymb{Q}}^{\pdag} c_{\bsymb{K}+\bsymb{Q}}^{\pdag} \\
        &\mbox{\quad}+ \sum_{j, \bsymb{K}} \left[ V_{\bsymb{G}_j}^{(\rm{m, e})} c_{\bsymb{K} + \bsymb{G}_j}^\dagger c_\bsymb{K}^{\pdag} - V_{\bsymb{G}_j}^{(\rm{m, h})} d_{\bsymb{K} + \bsymb{G}_j}^\dagger d_\bsymb{K}^{\pdag} \right] \\
        &= H_{\rm free} + H_{\rm e-h} + H_{\rm m}
\end{eqnarray*}
with the electron (hole) dispersion treated in the effective mass approximation $\epsilon_\bsymb{K}^{(\rm{e})} = \frac{\hbar^2 \bsymb{K}^2}{2m_{\rm e}} + E_{\rm gap}$ ($\epsilon_\bsymb{K}^{(\rm{h})} = \frac{\hbar^2 \bsymb{K}^2}{2m_{\rm h}}$), the energy of the band gap $E_{\rm gap}$, the effective mass of the electron (hole) $m_{\rm e}$ ($m_{\rm h}$) and the interaction potential $V_\bsymb{Q}$ between electrons and holes.

We start by introducing the \textit{homogeneous} exciton creation operator $Y_{l, \bsymb{\kappa}}^\dagger = \sum_\bsymb{K} \Phi_l(\bsymb{K} - \mu_{\rm e} \bsymb{\kappa}) c_\bsymb{K}^\dagger d_{\bsymb{\kappa} - \bsymb{K}}^\dagger$ with $\mu_{\rm e/h} = m_{\rm e/h} / M$ and total mass $M=m_{\rm e} + m_{\rm h}$, that creates an eigenstate of the Hamiltonian without the \moire potential, i.e., $H_{\rm free} + H_{\rm e-h}$, with center of mass wave vector $\bsymb{\kappa}$.
The corresponding annihilation operator is $Y_{l, \bsymb{\kappa}}^{\pdag}$.
Note that we have introduced the term \textit{homogeneous} exciton to distinguish it from the \textit{\moire} exciton that includes the \moire potential.
The one-exciton state $Y_{l, \bsymb{\kappa}}^\dagger \ket{0}$ satisfies the eigenvalue equation
\begin{eqnarray*}
    (H_{\rm free} + H_{\rm e-h}) Y_{l, \bsymb{\kappa}}^\dagger \ket{0} = E_{l, \bsymb{\kappa}} Y_{l, \bsymb{\kappa}}^\dagger \ket{0}
\end{eqnarray*}
with ground state (electron-hole vacuum) $\ket{0}$, eigenenergies $E_{l, \bsymb{\kappa}}$, and quantum number of the relative motion $l$.
This leads to a Wannier equation \cite{kira2011sem} for the coefficients $\Phi_l$, i.e., the exciton wave function in reciprocal space, according to
\begin{eqnarray*}
     \sum_\bsymb{Q} \left(\hbar^2 \frac{\bsymb{K}^2}{2 \mu} \delta_{\bsymb{Q}, \bsymb{0}} - V_{\bsymb{Q}} \right) \Phi_l(\bsymb{K} + \bsymb{Q}) = E_l \Phi_l(\bsymb{K})
\end{eqnarray*}
with reduced mass $\mu = m_{\rm e} m_{\rm h} / M$ and the energies of the homogeneous excitons
\begin{eqnarray}
    E_{l, \bsymb{\kappa}} = \frac{\hbar^2 \bsymb{\kappa}^2}{2M} + E_{\rm gap} + E_l.
    \label{eq:app_homogeneous_ex}
\end{eqnarray}

\begin{figure}[tb]
    \centering
    \includegraphics[width=0.65\linewidth]{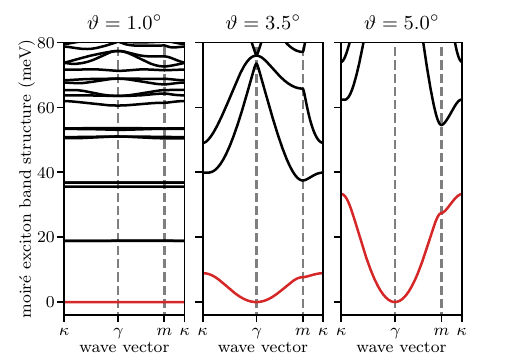}
    \caption{\Moire exciton dispersion relations for twist angles of $1^\circ$, $3.5^\circ$, and $5^\circ$ (from left to right).
        The lowest bands (red lines) have been used for the calculation of the polaritons.}
    \label{app:fig8}
\end{figure}

These homogeneous excitons additionally experience the \moire potential $H_{\rm m}$.
Assuming only pair excitations of electrons and holes and a small exciton density, the electronic transitions resulting from $H_{\rm m}$ can be approximated by \cite{katsch2018the}
\begin{numparts}
\begin{eqnarray}
\label{eq:app_e_Y}
    &c_{\bsymb{K} + \bsymb{G}_j}^\dagger  c_\bsymb{K}^{\pdag} \approx \sum_{\bsymb{K}'} c_{\bsymb{K} + \bsymb{G}_j}^\dagger d_{\bsymb{K}'}^\dagger d_{\bsymb{K}'}^{\pdag} c_\bsymb{K}^{\pdag} \\
    &=\sum_{l, l', \bsymb{\kappa}} \Phi_{l'}^*(\bsymb{K} - \mu_{\rm e} \bsymb{\kappa} +\mu_{\rm h}\bsymb{G}_j) \Phi_l(\bsymb{K} - \mu_{\rm e} \bsymb{Q}) Y_{l', \bsymb{\kappa} + \bsymb{G}_j}^\dagger Y_{l, \bsymb{\kappa}}^{\pdag},\nonumber\\
\label{eq:app_h_Y}
    &d_{\bsymb{K} + \bsymb{G}_j}^\dagger d_{\bsymb{K}}^{\pdag} \approx \sum_{\bsymb{K}'} c_{\bsymb{K}'}^\dagger d_{\bsymb{K} + \bsymb{G}_j}^\dagger d_{\bsymb{K}}^{\pdag} c_{\bsymb{K}'}^{\pdag} \\
    &=\sum_{l, l', \bsymb{\kappa}} \Phi_{l'}^*(\mu_{\rm h} \bsymb{\kappa} - \bsymb{K} -\mu_{\rm e}\bsymb{G}_j) \Phi_l(\mu_{\rm h} \bsymb{\kappa} - \bsymb{K}) Y_{l', \bsymb{\kappa} + \bsymb{G}_j}^\dagger Y_{l, \bsymb{\kappa}}^{\pdag}. \nonumber
\end{eqnarray}
\end{numparts}
With these approximations the Hamiltonian can be written in terms of homogeneous exciton operators
\begin{eqnarray}
    H = \sum_{l, \bsymb{\kappa}} E_{l, \bsymb{\kappa}}^{\pdag} Y_{l, \bsymb{\kappa}}^\dagger Y_{l, \bsymb{\kappa}}^{\pdag} + \sum_{{l,l', \bsymb{\kappa}, \bsymb{G}_j}} \tilde{V}_{\bsymb{G}_j}^{(l,l')} Y_{l', \bsymb{\kappa} + \bsymb{G}_j}^\dagger Y_{l, \bsymb{\kappa}}^{\pdag} 
    \label{eq:app_H_homogeneous_ex}
\end{eqnarray}
with 
\begin{eqnarray*}
    \tilde{V}_{\bsymb{G}}^{(l,l')} &= \left[ V_{\bsymb{G}}^{(\rm{m, e})} F_{l,l'}^{\pdag}(\mu_{\rm h} \bsymb{G}) - V_{\bsymb{G}}^{(\rm{m, h})} F_{l,l'}^{\pdag}(-\mu_{\rm e} \bsymb{G}) \right], \\
    F_{l,l'}(\bsymb{Q}) &= \sum_{\bsymb{K}} \Phi_{l'}^*(\bsymb{K} + \bsymb{Q}) \Phi_l^{\pdag}(\bsymb{K}),
\end{eqnarray*}
where $\tilde{V}_{\bsymb{G}}^{(l,l')}$ is a Fourier component of the \moire potential for excitons, resulting from the \moire potentials of electrons and holes and modified by the exciton form factor $F_{l,l'}(\bsymb{Q})$.
As seen in Eq.~\ref{eq:app_H_homogeneous_ex}, this \moire potential in general couples exciton states that differ by a reciprocal \moire lattice vector $\bsymb{G}_j$ both within the same exciton band ($l=l'$) and in different bands ($l \ne l'$).
In TMDCs the binding energy of the 1s exciton is typically very large \cite{berghauser2014ana,chernikov2014exc}, such that we can restrict ourselves to the 1s exciton with $l=l'=\rm{1s}$ and we neglect interband transitions that are induced by the \moire potential, thus assuming a single form factor $F(\bsymb{Q})$.
In this case the Hamiltonian can be further simplified to
\begin{eqnarray}
     H = \sum_{\bsymb{\kappa}} E_{\bsymb{\kappa}}^{\pdag} Y_{\bsymb{\kappa}}^\dagger Y_{\bsymb{\kappa}}^{\pdag} + \sum_{\bsymb{\kappa}, j} \tilde{V}_{\bsymb{G}_j}^{\pdag} Y_{\bsymb{\kappa} + \bsymb{G}_j}^\dagger Y_{\bsymb{\kappa}}^{\pdag},
     \label{eq:app_Hmoire}
\end{eqnarray}
where we dropped all band indices $l$, $l'$.
This interaction is of the same form as, for example, derived in Ref.~\cite{knorr2022exc} and describes an exciton in a periodic \moire potential with reciprocal lattice vectors $\bsymb{G}_j$.
Note that Eq.~\ref{eq:app_Hmoire} can also directly be interpreted as a homogeneous exciton coupled to a \moire potential for excitons [see Eq.~\ref{eq:moire_potential}]
\begin{eqnarray*}
    V(\bsymb{r}) = \sum_{j} \tilde{V}_{\bsymb{G}_j} e^{i\bsymb{G}_j \cdot \bsymb{r}}.
\end{eqnarray*}

As seen in Eq.~\ref{eq:app_Hmoire}, the \moire potential only couples excitons that differ by a reciprocal lattice vector of the \moire lattice $\bsymb{G}_j$.
Each wave vector $\bsymb{\kappa}$ can be uniquely written as a sum of a contribution within the 1st MBZ, $\bsymb{k} \in \mathcal{B}_{\rm m}$ and a reciprocal \moire lattice vector $\bsymb{G}_m$, where $m$ counts all reciprocal \moire lattice vectors.
Therefore, each sum $\sum_\bsymb{\kappa}$ can be uniquely replaced by $\sum_{m, \bsymb{k}}$ with $\bsymb{\kappa} = \bsymb{k} + \bsymb{G}_m$.
Then the Hamiltonian from Eq.~\ref{eq:app_Hmoire} can be diagonalized by introducing the \moire exciton creation and annihilation operators
\begin{numparts}
\begin{eqnarray}
    X_{n, \bsymb{k}}^\dagger &= \sum_{m} \varphi_{\bsymb{k}, \bsymb{G}_m}^{(n)} Y_{\bsymb{k} + \bsymb{G}_m}^\dagger,
    \label{eq:app_moire_ex_a_operator}
    \\
    X_{n, \bsymb{k}} &= \sum_{m} \left[\varphi_{\bsymb{k}, \bsymb{G}_m}^{(n)} \right]^* Y_{\bsymb{k} + \bsymb{G}_m},
    \label{eq:app_moire_ex_c_operator}
\end{eqnarray}
\end{numparts}
and solving the eigenvalue equation
\begin{eqnarray}
    \sum_s \left[ \left(E_{\bsymb{k} + \bsymb{G}_m} - \hbar \omega_{n, \bsymb{k}} \right) \delta_{s, m} + \tilde{V}_{\bsymb{G}_m-\bsymb{G}_s} \right] \varphi_{\bsymb{k}, \bsymb{G}_s}^{(n)} = 0
    \label{eq:app_eigenvalue_moire}
\end{eqnarray}
for each $\bsymb{k} \in \mathcal{B}_{\rm m}$, where  $E_{\bsymb{k} + \bsymb{G}_m}$ is the energy of the homogeneous exciton given by Eq.~\ref{eq:app_homogeneous_ex}.
This leads to the dispersion relation of the \moire exciton $\hbar \omega_{n, \bsymb{k}}$ and their coefficients $\varphi_{\bsymb{k}, \bsymb{G}_s}^{(n)}$.
The index $n$ acts as a band index and results from the back-folding of the exciton bands into the smaller MBZ.

In Fig.~\ref{app:fig8} the \moire exciton dispersion relations are plotted for the three twist angles $1^\circ$, $3.5^\circ$, and $5^\circ$.
Indeed, we find for the smallest twist angle a flat lowest exciton band, while with increasing twist angle the curvature and the width of the lowest exciton band increase.

\section{Derivation of the \moire exciton-phonon coupling}
\label{app:sec_phonon_coupling}
We now derive the \moire exciton-phonon interaction starting again from the electron-hole picture.
The standard electron/hole-phonon interaction Hamiltonian is given by
\begin{eqnarray*}
    H_{\rm{int}} &= \sum_{\bsymb{K}, \bsymb{Q}, j} \hbar \left[g_{j, \bsymb{Q}}^{(\rm{e})} c_{\bsymb{K} + \bsymb{Q}}^\dagger c_{\bsymb{K}}^{\pdag} - g_{j, \bsymb{Q}}^{(\rm{h})} d_{\bsymb{K} + \bsymb{Q}}^\dagger d_{\bsymb{K}}^{\pdag} \right] \nonumber \\
    &\hspace{1.5cm} \times \left(b_{j, \bsymb{Q}}^{\pdag} + b_{j, -\bsymb{Q}}^\dagger \right)
\end{eqnarray*}
with electron/hole-phonon coupling constant $g_{j, \bsymb{Q}}^{(\rm{e/h})}$.
Replacing the electron/hole operators by the homogeneous exciton operators according to  Eqs.~\ref{eq:app_e_Y} and~\ref{eq:app_h_Y} leads to
\begin{eqnarray*}
    H_{\rm{int}} &= \sum_{{l, l', j, \bsymb{\kappa}, \bsymb{Q}}} \hbar g_{j, \bsymb{Q}}^{(l,l')} Y_{l', \bsymb{\kappa} + \bsymb{Q}}^\dagger Y_{l, \bsymb{\kappa}}^{\pdag} \left(b_{j, \bsymb{Q}}^{\pdag} + b_{j, -\bsymb{Q}}^\dagger \right)
\end{eqnarray*}
with the homogeneous exciton-phonon coupling element $g_{j, \bsymb{Q}}^{(l,l')} = g_{j, \bsymb{Q}}^{(\rm{e})} F_{l,l'}(\mu_{\rm h} \bsymb{Q}) - g_{j, \bsymb{Q}}^{(\rm{h})} F_{l,l'}(-\mu_{\rm e}\bsymb{Q})$.
Again restricting the calculations to 1s excitons with $l = l' = \rm{1s}$ due to the large exciton binding energy in TMDCs, the interaction Hamiltonian can be further simplified to
\begin{eqnarray}
    &H_{\rm{int}} = \sum_{j, \bsymb{\kappa}, \bsymb{Q}} \hbar g_{j, \bsymb{Q}} Y_{\bsymb{\kappa} + \bsymb{Q}}^\dagger Y_{\bsymb{\kappa}}^{\pdag} \left(b_{j, \bsymb{Q}}^{\pdag} + b_{j, -\bsymb{Q}}^\dagger \right) \nonumber \\
    &= \sum_{{n, n', j, \bsymb{k}, \bsymb{Q}, \bsymb{G}}} \hbar \mathcal{G}_{j, \bsymb{k}, \bsymb{Q}}^{(n, n')} X_{n', \bsymb{k} + \bsymb{Q} + \bsymb{G}}^\dagger X_{n, \bsymb{k}}^{\pdag} \left(b_{j, \bsymb{Q}}^{\pdag} + b_{j, -\bsymb{Q}}^\dagger \right),
    \label{eq:app_H_ex-phon}
\end{eqnarray}
where we have used the \moire exciton operators $X_{n, \bsymb{k}}^{\pdag}$ ($X_{n, \bsymb{k}}^\dagger$) defined in Eqs.~\ref{eq:app_moire_ex_a_operator} and~\ref{eq:app_moire_ex_c_operator} and introduced the \moire exciton-phonon coupling constant
\begin{eqnarray*}
    \mathcal{G}_{j, \bsymb{k}, \bsymb{Q}}^{(n, n')} &= g_{j, \bsymb{Q}} f_{\bsymb{k}, \bsymb{Q}}^{(n,n')} \\
    &= \left[g_{j, \bsymb{Q}}^{(\rm{e})} F(\mu_{\rm h} \bsymb{Q}) - g_{j, \bsymb{Q}}^{(\rm{h})} F(-\mu_{\rm e}\bsymb{Q}) \right]
        f_{\bsymb{k}, \bsymb{Q}}^{(n,n')} 
\end{eqnarray*}
with the \moire form factor
\begin{eqnarray*}
    f_{\bsymb{k}, \bsymb{Q}}^{(n,n')} &= \sum_{m, \bsymb{G}} \left[\varphi_{\bsymb{k} + \bsymb{Q} + \bsymb{G}, \bsymb{G}_m - \bsymb{G}}^{(n')}\right]^* \varphi_{\bsymb{k}, \bsymb{G}_m}^{(n)}.
\end{eqnarray*}
Note, that $\bsymb{G}$ describes umklapp processes and is chosen such that $\bsymb{k} + \bsymb{Q} + \bsymb{G} \in \mathcal{B}_{\rm m}$.
In general the \moire form factor $f_{\bsymb{k}, \bsymb{Q}}^{(n,n')}$ depends on the twist angle, leading to a twist-angle dependent \moire exciton-phonon coupling.

\section{Perturbative Solution}
\label{app:sec_pert}

In the long-time limit the phonon part $\Gamma_{\Lambda, \Lambda'}(\bsymb{k}, t)$  in the equation of motion (\ref{eq:tcl}) reduces to Markovian, time-independent coefficients $\Gamma_{\Lambda, \Lambda'}^{\rm {Markov}}(\bsymb{k})$ [see Eq.~\ref{eq:Markov}].
This Markovian limit can be used as a starting point for a perturbative solution of Eq.~\ref{eq:tcl}.
For this purpose we separate the phonon part into the Markovian part according to Eq.~\ref{eq:Markov} and a time-dependent, non-Markovian correction $\delta \Gamma_{\Lambda, \Lambda'}(\bsymb{k}, t)$ according to Eq.~\ref{eq:separation}.

To determine the non-Markovian part we note that in Eq.~\ref{eq:Gamma} we can split off an interval $\left[-\epsilon, \epsilon \right]$ around $\Omega = 0$ with $\epsilon > 0$, 
\begin{eqnarray*}
    \Gamma_{\Lambda, \Lambda'}(\bsymb{k}, t) &= \int\limits_{-\infty}^\infty \dd{\Omega} \rho_{\Lambda, \Lambda'}(\bsymb{k}, \Omega) \int_{0}^{t} \dd{\tau} e^{-i\Omega \tau}
    \\
    &= \int\limits_{-\infty}^\infty \dd{\Omega} \rho_{\Lambda, \Lambda'}(\bsymb{k}, \Omega) \frac{e^{-i\Omega t}-1}{-i\Omega}
    \\
    &= \lim\limits_{\epsilon \rightarrow 0} {\Bigg{[}}
    \int\limits_{\mathbb{R} \setminus \left[-\epsilon, \epsilon \right]} \dd{\Omega} \rho_{\Lambda, \Lambda'}(\bsymb{k}, \Omega) \frac{e^{-i\Omega t}-1}{-i\Omega}
    \\
    &+ \int\limits_{-\epsilon}^{\epsilon} \dd{\Omega} \rho_{\Lambda, \Lambda'}(\bsymb{k}, \Omega) \frac{e^{-i\Omega t}-1}{-i\Omega}
    {\Bigg{]}}.
\end{eqnarray*}
The first integral on the right hand side can by definition be written as a principal value integral $\pv \int \dd{\Omega}$.
Since  $\lim\limits_{\Omega \rightarrow 0} \frac{e^{-i\Omega t}-1}{-i\Omega} = t$, the integrand of the second integral is finite in a region around $\Omega = 0$ for any finite time $t$ if $\rho_{\Lambda, \Lambda'}(\bsymb{k}, \Omega)$ has a removable singularity.
Therefore, the whole integral vanishes in the limit $\epsilon \rightarrow 0$.
Then we can further split the principal value integral into two parts and get
\begin{eqnarray*}
    \Gamma_{\Lambda, \Lambda'}(\bsymb{k}) =& -i\pv \int \dd{\Omega} \frac{\rho_{\Lambda, \Lambda'}(\bsymb{k}, \Omega)}{\Omega} \nonumber \\
    &+ i \pv \int \dd{\Omega} \frac{\rho_{\Lambda, \Lambda'}(\bsymb{k}, \Omega)}{\Omega} e^{-i\Omega t}.
\end{eqnarray*}
The first term agrees with the imaginary part of the Markovian limit of $\Gamma_{\Lambda, \Lambda'}$ [see Eq.~\ref{eq:ImMarkov}].
Using Eqs~\ref{eq:Markov} and \ref{eq:separation}, we thus obtain for the non-Markovian part
\begin{eqnarray*}
    \delta \Gamma_{\Lambda, \Lambda'}(\bsymb{k}, t) &= i \pv \int \dd{\Omega} \frac{\rho_{\Lambda, \Lambda'}(\bsymb{k}, \Omega)}{\Omega} e^{-i\Omega t} - \pi  \rho_{\Lambda, \Lambda'}(\bsymb{k}, 0).
 \end{eqnarray*}
Based on this separation, we obtain the equation of motion (\ref{eq:pol_split})
\begin{eqnarray*}
    \dv{t} \tilde{P}_{\Lambda, \bsymb{k}}(t) &= \sum_{\Lambda'} \left[ -i A_{\Lambda, \Lambda'}(\bsymb{k}) - \delta \Gamma_{\Lambda, \Lambda'}(\bsymb{k}, t) \right] \tilde{P}_{\Lambda', \bsymb{k}}(t)
\end{eqnarray*}
with the bare polariton and the Markovian part included in the matrix
\begin{eqnarray*}
    A_{\Lambda, \Lambda'}(\bsymb{k}) &= \Delta_{\Lambda, \bsymb{k}} \delta_{\Lambda, \Lambda'} - i \Gamma_{\Lambda, \Lambda'}^{\rm{Markov}}(\bsymb{k}).
\end{eqnarray*} 
A rotation, which diagonalizes $(A_{\Lambda, \Lambda'})$ according to $D_{\Lambda}(\bsymb{k}) = \sum_{\Lambda', \Lambda''} R_{\Lambda, \Lambda'}^{-1}(\bsymb{k}) \penalty 0 A_{\Lambda', \Lambda''}(\bsymb{k}) \penalty 0 R_{\Lambda'', \Lambda}(\bsymb{k})$ with $D_{\Lambda}(\bsymb{k})$ containing the eigenvalues of $A_{\Lambda, \Lambda'}(\bsymb{k})$ leads to the differential equation
\begin{eqnarray}
    \dv{t} x_{\Lambda, \bsymb{k}}(t) = &- i D_{\Lambda}(\bsymb{k}) x_{\Lambda, \bsymb{k}}(t) \nonumber \\
    &- \sum_{\Lambda'} \delta \Gamma'_{\Lambda, \Lambda'}(\bsymb{k}, t) x_{\Lambda', \bsymb{k}}(t)
    \label{app:pert_dgl_diag}
\end{eqnarray}
with $x_{\Lambda, \bsymb{k}}(t) = \sum_{\Lambda'} R_{\Lambda, \Lambda'}^{-1}(\bsymb{k}) \tilde{P}_{\Lambda', \bsymb{k}}(t)$ and $\delta \Gamma'_{\Lambda, \Lambda'}(\bsymb{k}, t) = \sum_{\Lambda'', \Lambda'''} R^{-1}_{\Lambda, \Lambda''}(\bsymb{k}) \penalty 0 \delta \Gamma_{\Lambda'', \Lambda'''}(\bsymb{k}, t) \penalty 0 R_{\Lambda''', \Lambda'}(\bsymb{k})$.
This diagonalization is guaranteed to work as long as $A_{\Lambda, \Lambda'}(\bsymb{k})$ has pairwise distinct eigenvalues \cite{fox1964int}.
Since in general $A_{\Lambda, \Lambda'}$ is not Hermitian or symmetric, the eigenvalues will be complex.
Due to the structure of the equation, we can identify these eigenvalues $D_{\Lambda}(\bsymb{k})$ with the peak positions (real parts) and line widths (imaginary parts) of the polariton, such that the diagonalization is possible as long as the polaritons have distinct energies.

Via the Laplace transformation
\begin{eqnarray}
    \tilde{x}_{\Lambda, \bsymb{k}}(s) = \int\limits_0^\infty \dd{t} x_{\Lambda, \bsymb{k}}(t) e^{-st}
    \label{app:Laplace}
\end{eqnarray}
the differential equation from \ref{app:pert_dgl_diag} can be converted into a Fredholm equation \cite{courant1953met}
\begin{eqnarray}
    \bsymb{\tilde{x}}_\bsymb{k}(s) = \bsymb{f}_\bsymb{k}(s) + \left(\mathcal{K}_\bsymb{k} \bsymb{\tilde{x}}_\bsymb{k} \right)(s)
    \label{app:fredholm}
\end{eqnarray}
with vectors $\bsymb{\tilde{x}}_\bsymb{k} = \left(\tilde{x}_{0, \bsymb{k}}, \dots, \tilde{x}_{N, \bsymb{k}} \right)^T$ and $\bsymb{f}_\bsymb{k} = \left(f_{0, \bsymb{k}}, \dots, f_{N, \bsymb{k}} \right)^T$ with components
\begin{eqnarray}
    f_{\Lambda, \bsymb{k}}(s) = \frac{x_{\Lambda, \bsymb{k}}(0)}{s + i D_{\Lambda}(\bsymb{k})},
    \label{app:eq_f}
\end{eqnarray}
the operator $\mathcal{K}_{\bsymb{k}}$ defined by 
\begin{eqnarray}
    \left[\left(\mathcal{K}_{\bsymb{k}} \bsymb{\tilde{x}} \right)(s) \right]_\Lambda &= \sum_{\Lambda'}
        \bigg{[}
             \frac{\pi \rho_{\Lambda, \Lambda'}'(\bsymb{k}, 0) }{s + iD_{\Lambda}(\bsymb{k})} \tilde{x}_{\Lambda', \bsymb{k}}(s)
             \nonumber \\
        &-i \pv \int \dd{\Omega} \frac{\rho'_{\Lambda, \Lambda'}(\bsymb{k}, \Omega) }{\Omega \left[s + i D_{\Lambda}(\bsymb{k}) \right]} \tilde{x}_{\Lambda', \bsymb{k}}(s + i\Omega) 
        \bigg{]}
    \nonumber
    \\
    & = \frac{f_{\Lambda, \bsymb{k}}(s)}{x_{\Lambda, \bsymb{k}}(0)} \sum_{\Lambda'} 
        \bigg{[}
            \pi \rho'_{\Lambda, \Lambda'}(\bsymb{k}, 0) \tilde{x}_{\Lambda', \bsymb{k}}(s)
        \nonumber \\
            &-i \pv \int \dd{\Omega} \frac{\rho'_{\Lambda, \Lambda'}(\bsymb{k}, \Omega)}{\Omega} \tilde{x}_{\Lambda', \bsymb{k}}(s + i\Omega)
        \bigg{]}
    \label{app:pert_operator}
\end{eqnarray}
and the rotated gPSD $\rho'_{\Lambda, \Lambda'}(\bsymb{k}, \Omega) = \sum_{\Lambda'', \Lambda'''} R^{-1}_{\Lambda, \Lambda''}(\bsymb{k}) \rho_{\Lambda'', \Lambda'''}(\bsymb{k}, \Omega) R_{\Lambda''', \Lambda'}(\bsymb{k})$.
Note that the Markovian part of the system is completely included in $\bsymb{f}_\bsymb{k}$ and the non-Markovian part in the operator $\mathcal{K}_{\bsymb{k}}$.
Assuming that $\mathcal{K}_\bsymb{k} \rightarrow \lambda \mathcal{K}_\bsymb{k}$ is a small perturbation, \ref{app:fredholm} can be solved via the power series \cite{courant1953met} $\tilde{\bsymb{x}}_\bsymb{k}(s) = \sum_{n} \lambda^n \bsymb{\Phi}_\bsymb{k}^{(n)}(s)$ with
\begin{eqnarray}
    \bsymb{\Phi}_\bsymb{k}^{(0)}(s) &= \bsymb{f}_\bsymb{k}(s) 
    \label{app:pert_iteration0}
    \\
    \bsymb{\Phi}_\bsymb{k}^{(n)}(s) &= \left[\mathcal{K}_\bsymb{k} \bsymb{\Phi}_\bsymb{k}^{(n-1)} \right](s)
    \label{app:pert_iterationn}
\end{eqnarray}
which is an iterative solution since $\bsymb{\Phi}_\bsymb{k}^{(n)}(s)$ depends on the next lower order $\bsymb{\Phi}_\bsymb{k}^{(n-1)}$.
Both Eqs.~\ref{eq:tcl} and \ref{app:fredholm} describe the polarization of the polaritons with wave vector $\bsymb{k}$ including the influence of phonons, taking the polariton and phonon dispersion and the coupling strength to the phonons into account.
These quantities are all hidden in the gPSD $\rho_{\Lambda, \Lambda'}(\bsymb{k}, \Omega)$ and the rotated gPSD $\rho'_{\Lambda, \Lambda'}(\bsymb{k}, \Omega)$, respectively.
The optical absorption spectrum (Eq.~\ref{eq:spectrum}) is then given by a superposition of all $\Phi_{\Lambda, \bsymb{k}}^{(n)}(s = -i\omega)$
\begin{eqnarray*}
    \alpha_{\bsymb{k}}(\omega) &= \Re\left[\sum_{\Lambda} \conj{\beta_{\bsymb{k}; \Lambda, 0}} \int\limits_0^\infty \dd{t} \tilde{P}_{\Lambda, \bsymb{k}}(t) e^{i\omega t} \right] \nonumber \\
    &= \Re\left[\sum_{\Lambda, \Lambda', n} \conj{\beta_{\bsymb{k}; \Lambda, 0}} R_{\Lambda, \Lambda'}(\bsymb{k}) \lambda^n \Phi_{\Lambda', \bsymb{k}}^{(n)}\left(-i\omega \right) \right].
\end{eqnarray*}
We arrive at an approximate solution by only taking a finite number of $\Phi_{\Lambda, \bsymb{k}}^{(n)}(s)$ into account.
Up to 1st order it reads
%\begin{widetext}
    \begin{eqnarray}
        \Phi_{\Lambda, \bsymb{k}}^{(0)}(s) + \Phi_{\Lambda, \bsymb{k}}^{(1)}(s) 
        =& f_{\Lambda, \bsymb{k}}(s) + \sum_{\Lambda'} \left[
            \frac{\pi \rho'_{\Lambda, \Lambda'}(\bsymb{k}, 0)}{s+iD_\Lambda} f_{\Lambda', \bsymb{k}}(s)
            \right.
            \nonumber \\
            &\left.
            -i\pv \int \dd{\Omega} 
        \frac{\rho'_{\Lambda, \Lambda'}(\bsymb{k}, \Omega)}{\Omega\left[s+iD_{\Lambda}(\bsymb{k}) \right]}
        f_{\Lambda', \bsymb{k}}(s+i\Omega)\right].
        \label{app:pert_iteration2}
    \end{eqnarray}
%\end{widetext}
The first term in \ref{app:pert_iteration2} includes the Markovian part of the phonon interaction and leads to peaks at $s = -iD_{\Lambda}(\bsymb{k})$, which depend on the temperature.
The second term includes the non-Markovian part that also gives a contribution to the polariton peaks via the terms $\sim \frac{1}{s + iD_{\Lambda}(\bsymb{k})}$ and also an additional term that depends on the shape of the rotated gPSD $\int\dd{\Omega} \frac{\rho'_{\Lambda, \Lambda'}(\bsymb{k}, \Omega)}{\Omega \left[s + i\Omega + i D_{\Lambda'}(\bsymb{k})\right]}$.
By inserting $f_{\Lambda', \bsymb{k}}(s+i\Omega)$ from \ref{app:eq_f} into the latter term, one can see that it leads to a strong contribution at $s = -i\left[\Omega + D_{\Lambda'}(\bsymb{k}) \right]$ weighted by the gPSD at $\Omega$, resulting in phonon sidebands.

Assuming that the gPSD has localized maxima at $\Omega_j$ these would narrow  to additional peaks at $s = -i\left[\Omega_j + D_{\Lambda'}(\bsymb{k}) \right]$ after this first iteration step.
Notably, if $\Omega_j \approx - D_{\Lambda'}(\bsymb{k})$ this leads to additional peaks at $s \approx 0$.
GPSD contributions at $\Omega = 0$ would lead to increased corrections at one of the polariton peaks.
These features are discussed in Sec.~\ref{sec:comp_pert}.

\section{Twist-angle dependencies of additional quantities}\label{app:twist}
In Sec.~\ref{sec:twist_dep} we investigated the twist-angle dependence of the UP and LP peak position and width and discussed the general shape of these curves.
We found a twist angle region with a significantly increased energy shift and broadening of the UP, which we identified as the transition region.
We traced these effects back to energy-conserving phonon-assisted transitions to the van Hove singularities.
In this section we present the twist-angle dependencies of further quantities that show the same transition between small and large twist angles.
\begin{figure}[tb]
    \centering
    \includegraphics[width=1.0\linewidth]{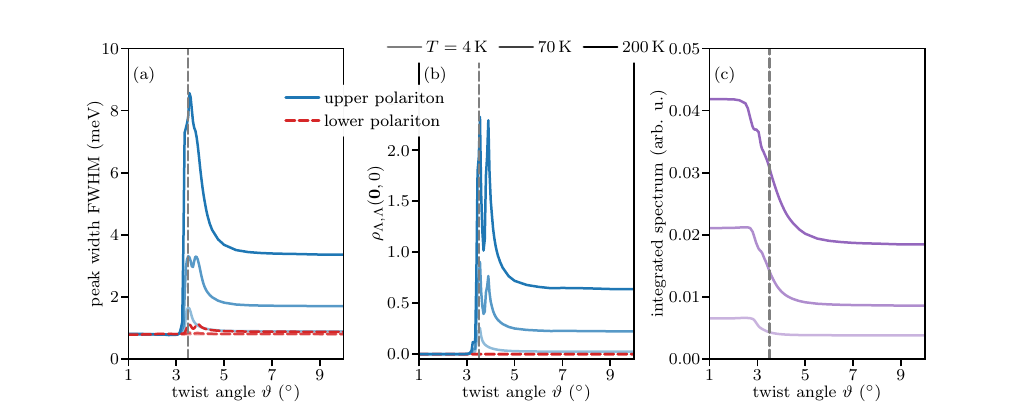}
    \caption{Twist angle scan of the peak width (a) (identical to Fig.~\ref{fig:fig7} (a)) and $\rho_{\Lambda, \Lambda}(\bsymb{0}, \Omega)$ at $\Omega = 0$ of the UP and LP peaks for different temperatures. (c) Twist angle scan of the phonon sideband-integrated spectrum for different temperatures. The twist angle $\vartheta = 3.5^\circ$ is marked with a vertical dashed line.}
    \label{app:fig9}
\end{figure}

For reference, Fig.~\ref{app:fig9}(a) shows the peak width of the UP and LP (same as Fig.~\ref{fig:fig7}(a)).
In addition, Fig.~\ref{app:fig9}(b) depicts the gPSD value $\rho_{\Lambda, \Lambda}(\boldsymbol{0}, \Omega)$ at $\Omega = 0$, and (c) the absorption spectrum integrated from $-5\,$meV to $+5\,$meV, i.e., across the isolated phonon sidebands, for different temperatures.
The gPSD at $\Omega = 0$ (b) directly reflects energy-conserving phonon transitions and is the main contribution of the gPSD to the broadening of the polariton peaks.
Therefore, we find a similar trend as in (a) with a slightly better resolution of the double peak structure in the transition region.
These peaks can be directly traced back to phonon-assisted transitions to van Hove singularities in the polariton dispersion, where one peak corresponds to phonon absorption and the other to phonon emission processes.
According to Eqs.~(\ref{app:pert_dgl_diag}) and (\ref{app:fredholm}) the gPSD contributes to the spectrum in multiple ways, namely, its value at $\Omega = 0$, the area under the curve $\rho_{\Lambda, \Lambda'}(\bsymb{0}, \Omega) / \Omega$ and a convolution with this function (see \ref{app:sec_pert}).
This directly explains the slight differences between Figs.~\ref{app:fig9}(a) and (b).

For small twist angles we find the emergence of phonon sidebands in the spectra in the energy range around $\hbar \omega = 0$ (see Fig.~\ref{fig:fig4}(c)).
Otherwise, for large twist angles, the sidebands continuously range from the UP peak to lower energies (see Fig.~\ref{fig:fig4}(a)).
This transition in the character of phonon sidebands illustrates that the spectral behavior is dominated by energy-conserving transitions at large twist angles and by the existence of a  polariton band gap at small twist angles.
Figure~\ref{app:fig9}(c) shows the integrated spectrum around the phonon sidebands in the center between UP and LP peak.
We choose the energy range from $-5\,$meV to $+5\,$meV, i.e.,
\begin{eqnarray}
    \alpha_{\bsymb{k}}^{\rm {(int)}} = \int\limits_{-5\,\rm {meV}}^{+5\,\rm {meV}} \dd{\omega} \alpha_{\bsymb{k}}(\omega),
\end{eqnarray}
such that this quantity contains all spectral contributions close to $\hbar \omega = 0$, but not the main polariton peaks.
While at large twist angles the only contribution is given by a fraction of the continuous phonon sideband ranging from the UP peak to lower energies resulting in the plateau at smaller values, at small twist angles it includes the whole phonon sideband leading to the plateau at larger values (see Fig.~\ref{app:fig9}(c)).
We again find that the transition between these two flat regions happens around $\vartheta \approx 3.5^\circ$.
Contrary to (a) and (b), where we find sharp features (peaks) exactly at that twist angle, the transition in (c) is smoother.
This supports our findings, that the energy-conserving transitions mainly contribute to the UP peak, while the phonon sidebands at $\hbar \omega \approx 0$ result from the formation of the polariton band gap.

\section*{References}
%\bibliographystyle{iopart-num}
%\bibliography{lit}

\begin{thebibliography}{10}
    \expandafter\ifx\csname url\endcsname\relax
    \def\url#1{{\tt #1}}\fi
    \expandafter\ifx\csname urlprefix\endcsname\relax\def\urlprefix{URL }\fi
    \providecommand{\eprint}[2][]{\url{#2}}
    
    \bibitem{he2021moi}
    He F, Zhou Y, Ye Z, Cho S~H, Jeong J, Meng X and Wang Y 2021 {\em ACS Nano\/}
    {\bf 15} 5944--5958
    
    \bibitem{yoo2019ato}
    Yoo H, Engelke R, Carr S, Fang S, Zhang K, Cazeaux P, Sung S~H, Hovden R, Tsen
    A~W, Taniguchi T, Watanabe K, Yi G~C, Kim M, Luskin M, Tadmor E~B, Kaxiras E
    and Kim P 2019 {\em Nat. Mater.\/} {\bf 18} 448--453
    
    \bibitem{wang2020corr}
    Wang L, Shih E~M, Ghiotto A, Xian L, Rhodes D~A, Tan C, Claassen M, Kennes D~M,
    Bai Y, Kim B {\em et~al.\/} 2020 {\em Nat. Mater.\/} {\bf 19} 861--866
    
    \bibitem{zhang2018moi}
    Zhang N, Surrente A, Baranowski M, Maude D~K, Gant P, Castellanos-Gomez A and
    Plochocka P 2018 {\em Nano Lett.\/} {\bf 18} 7651--7657
    
    \bibitem{villafane2023twi}
    Villafa\~ne V, Kremser M, H\"ubner R, Petri{\'e} M~M, Wilson N~P, Stier A~V,
    M\"uller K, Florian M, Steinhoff A and Finley J~J 2023 {\em Phys. Rev.
        Lett.\/} {\bf 130}(2) 026901
    
    \bibitem{knorr2022exc}
    Knorr W, Brem S, Meneghini G and Malic E 2022 {\em Phys. Rev. Mater.\/} {\bf
        6}(12) 124002
    
    \bibitem{brem2024opt}
    Brem S and Malic E 2024 {\em 2D Mater.\/} {\bf 11} 025030
    
    \bibitem{wu2022evi}
    Wu B, Zheng H, Li S, Ding J, He J, Zeng Y, Chen K, Liu Z, Chen S, Pan A and Liu
    Y 2022 {\em Light Sci. Appl.\/} {\bf 11} 166
    
    \bibitem{wang2021moi}
    Wang X, Zhu J, Seyler K~L, Rivera P, Zheng H, Wang Y, He M, Taniguchi T,
    Watanabe K, Yan J {\em et~al.\/} 2021 {\em Nat. Nanotechnol.\/} {\bf 16}
    1208--1213
    
    \bibitem{seyler2019sig}
    Seyler K~L, Rivera P, Yu H, Wilson N~P, Ray E~L, Mandrus D~G, Yan J, Yao W and
    Xu X 2019 {\em Nature\/} {\bf 567} 66--70
    
    \bibitem{li2023rev}
    Li Z, Lai J~M and Zhang J 2023 {\em J. Semicond.\/} {\bf 44} 011902
    
    \bibitem{koshino2019moi}
    Koshino M and Son Y~W 2019 {\em Phys. Rev. B\/} {\bf 100}(7) 075416
    
    \bibitem{koshino2020eff}
    Koshino M and Nam N~N~T 2020 {\em Phys. Rev. B\/} {\bf 101}(19) 195425
    
    \bibitem{lin2018moi}
    Lin M~L, Tan Q~H, Wu J~B, Chen X~S, Wang J~H, Pan Y~H, Zhang X, Cong X, Zhang
    J, Ji W, Hu P~A, Liu K~H and Tan P~H 2018 {\em ACS Nano\/} {\bf 12}
    8770--8780
    
    \bibitem{nam2017lat}
    Nam N~N~T and Koshino M 2017 {\em Phys. Rev. B\/} {\bf 96}(7) 075311
    
    \bibitem{chuang2022eme}
    Chuang H~J, Phillips M, McCreary K~M, Wickramaratne D, Rosenberger M~R, Oleshko
    V~P, Proscia N~V, Lohmann M, O’Hara D~J, Cunningham P~D, Hellberg C~S and
    Jonker B~T 2022 {\em ACS Nano\/} {\bf 16} 16260--16270
    
    \bibitem{baek2020highly}
    Baek H, Brotons-Gisbert M, Koong Z, Campbell A, Rambach M, Watanabe K,
    Taniguchi T and Gerardot B~D 2020 {\em Sci. Adv.\/} {\bf 6} eaba8526
    
    \bibitem{wu2018the}
    Wu F, Lovorn T and MacDonald A~H 2018 {\em Phys. Rev. B\/} {\bf 97}(3) 035306
    
    \bibitem{yu2015ano}
    Yu H, Wang Y, Tong Q, Xu X and Yao W 2015 {\em Phys. Rev. Lett.\/} {\bf
        115}(18) 187002
    
    \bibitem{liu2023exa}
    Liu D, Zeng J, Jiang X, Tang L and Chen K 2023 {\em Phys. Rev. B\/} {\bf
        107}(8) L081402
    
    \bibitem{shabani2021dee}
    Shabani S, Halbertal D, Wu W, Chen M, Liu S, Hone J, Yao W, Basov D~N, Zhu X
    and Pasupathy A~N 2021 {\em Nat. Phys.\/} {\bf 17} 720--725
    
    \bibitem{lin2023rem}
    Lin B~H, Chao Y~C, Hsieh I~T, Chuu C~P, Lee C~J, Chu F~H, Lu L~S, Hsu W~T, Pao
    C~W, Shih C~K, Su J~J and Chang W~H 2023 {\em Nano Lett.\/} {\bf 23}
    1306--1312
    
    \bibitem{brem2020tun}
    Brem S, Linder{\"a}lv C, Erhart P and Malic E 2020 {\em Nano Lett.\/} {\bf 20}
    8534--8540
    
    \bibitem{gotting2022moi}
    G\"otting N, Lohof F and Gies C 2022 {\em Phys. Rev. B\/} {\bf 105}(16) 165419
    
    \bibitem{yu2017moi}
    Yu H, Liu G, Tang J, Xu X and Yao W 2017 {\em Sci. Adv.\/} {\bf 3} e1701696
    
    \bibitem{lohof2023con}
    Lohof F, Michl J, Steinhoff A, Han B, von Helversen M, Tongay S, Watanabe K,
    Taniguchi T, Höfling S, Reitzenstein S, Anton-Solanas C, Gies C and
    Schneider C 2023 {\em 2D Mater.\/} {\bf 10} 034001
    
    \bibitem{shinokita2021res}
    Shinokita K, Miyauchi Y, Watanabe K, Taniguchi T and Matsuda K 2021 {\em Nano
        Lett.\/} {\bf 21} 5938--5944
    
    \bibitem{lim2023mod}
    Lim S~Y, Kim H, Choi Y~W, Taniguchi T, Watanabe K, Choi H~J and Cheong H 2023
    {\em ACS Nano\/} {\bf 17} 13938--13947 pMID: 37410957
    
    \bibitem{fitzgerald2022twi}
    Fitzgerald J~M, Thompson J~J~P and Malic E 2022 {\em Nano Lett.\/} {\bf 22}
    4468--4474
    
    \bibitem{zhang2021van}
    Zhang L, Wu F, Hou S, Zhang Z, Chou Y~H, Watanabe K, Taniguchi T, Forrest S~R
    and Deng H 2021 {\em Nature\/} {\bf 591} 61--65
    
    \bibitem{forg2019cav}
    F{\"o}rg M, Colombier L, Patel R~K, Lindlau J, Mohite A~D, Yamaguchi H, Glazov
    M~M, Hunger D and H{\"o}gele A 2019 {\em Nat. Commun.\/} {\bf 10} 3697
    
    \bibitem{kasprzak2006bos}
    Kasprzak J, Richard M, Kundermann S, Baas A, Jeambrun P, Keeling J~M~J,
    Marchetti F~M, Szyma{\'n}ska M~H, Andr{\'e} R, Staehli J~L, Savona V,
    Littlewood P~B, Deveaud B and Dang L~S 2006 {\em Nature\/} {\bf 443} 409--414
    
    \bibitem{kessler2008lig}
    Kessler E, Grochol M and Piermarocchi C 2008 {\em Phy. Rev. B\/} {\bf 77}(8)
    085306
    
    \bibitem{rossi2002the}
    Rossi F and Kuhn T 2002 {\em Rev. Mod. Phys.\/} {\bf 74}(3) 895--950
    
    \bibitem{lengers2020the}
    Lengers F, Kuhn T and Reiter D~E 2020 {\em Phys. Rev. B\/} {\bf 101}(15) 155304
    
    \bibitem{lengers2021phon}
    Lengers F, Kuhn T and Reiter D~E 2021 {\em Phys. Rev. B\/} {\bf 104}(24)
    L241301
    
    \bibitem{richter2010tim}
    Richter M and Knorr A 2010 {\em Ann. Phys.\/} {\bf 325} 711--747
    
    \bibitem{ferreira2023sig}
    Ferreira B, Rosati R, Fitzgerald J~M and Malic E 2022 {\em 2D Mater.\/} {\bf
        10} 015012
    
    \bibitem{breuer2002the}
    Breuer H~P and Petruccione F 2002 {\em The theory of open quantum systems\/}
    (Oxford: Oxford University Press)
    
    \bibitem{gao2022sym}
    Gao Q and Khalaf E 2022 {\em Phys. Rev. B\/} {\bf 106}(7) 075420
    
    \bibitem{rupp2023ima}
    Rupp A, G{\"o}ser J, Li Z, Bilgin I, Baimuratov A and H{\"o}gele A 2023 {\em 2D
        Materials\/} {\bf 10} 045028
    
    \bibitem{lundt2016roo}
    Lundt N, Klembt S, Cherotchenko E, Betzold S, Iff O, Nalitov A~V, Klaas M,
    Dietrich C~P, Kavokin A~V, H{\"o}fling S and Schneider C 2016 {\em Nat.
        Commun.\/} {\bf 7} 13328
    
    \bibitem{jin2014int}
    Jin Z, Li X, Mullen J~T and Kim K~W 2014 {\em Phys. Rev. B\/} {\bf 90}(4)
    045422
    
    \bibitem{courant1953met}
    Courant R and Hilbert D 1953 {\em Methods of mathematical physics / 1\/} 1st ed
    (Wiley) ISBN 047017952X
    
    \bibitem{kira2011sem}
    Kira M and Koch S~W 2011 {\em Semiconductor quantum optics\/} (Cambridge
    University Press)
    
    \bibitem{katsch2018the}
    Katsch F, Selig M, Carmele A and Knorr A 2018 {\em Phys. Status Solidi B\/}
    {\bf 255} 1800185
    
    \bibitem{berghauser2014ana}
    Bergh\"auser G and Malic E 2014 {\em Phys. Rev. B\/} {\bf 89}(12) 125309
    
    \bibitem{chernikov2014exc}
    Chernikov A, Berkelbach T~C, Hill H~M, Rigosi A, Li Y, Aslan B, Reichman D~R,
    Hybertsen M~S and Heinz T~F 2014 {\em Phys. Rev. Lett.\/} {\bf 113}(7) 076802
    
    \bibitem{fox1964int}
    Fox L 1964 {\em An Introduction to numerical linear algebra\/} Monographs on
    numerical analysis (Oxford: Clarendon Pr.)
    
\end{thebibliography}
\providecommand{\newblock}{}

\end{document}